\documentclass[aps,prd,10pt,notitlepage,nofootinbib,superscriptaddress,showkeys,showpacs]{revtex4-1}
\usepackage{amsmath,amssymb,amsthm,latexsym,bbm,calc}
\usepackage[english]{babel}
\usepackage{graphicx,color}
\usepackage{xspace}
\usepackage{graphicx}
\usepackage{pifont,dsfont}
\usepackage{marvosym}
\usepackage{slashed}
\usepackage{subfigure}

\newcommand{\be}{\begin{equation}}
\newcommand{\ee}{\end{equation}}
\newcommand{\bqa}{\begin{eqnarray}}
\newcommand{\eqa}{\end{eqnarray}}
\newcommand{\bea}{\begin{eqnarray}}
\newcommand{\eea}{\end{eqnarray}}

\newcommand{\N}{\mathds{N}}

\newcommand{\R}{\mathds{R}}

\renewcommand{\N}{\mathbbm{N}}
\renewcommand{\R}{\mathbbm{R}}

\newcommand{\Z}{\mathbbm{Z}}
\newcommand{\alg}{{\mathfrak  g}}

\newcommand{\su}{\mathfrak{su}}
\renewcommand{\u}{\mathfrak{u}}

\newcommand{\cF}{{\cal F}}

\DeclareMathOperator{\Ad}{Ad}
\DeclareMathOperator{\id}{id}
\DeclareMathOperator{\tr}{tr}

\DeclareMathOperator{\SU}{SU}
\DeclareMathOperator{\U}{U}
\DeclareMathOperator{\Li}{Li}
\DeclareMathOperator{\rk}{rk}

\begin{document}

\title{\Large \bf Bubble divergences and gauge symmetries in spin foams}

\author{{\bf Valentin Bonzom}}\email{vbonzom@perimeterinstitute.ca}
\author{{\bf Bianca Dittrich}}\email{bdittrich@perimeterinstitute.ca}
\affiliation{Perimeter Institute for Theoretical Physics, 31 Caroline St. N, ON N2L 2Y5, Waterloo, Canada}

\date{\small\today}

\begin{abstract}
\noindent
 Spin foams are candidate state-sum models for transition amplitudes in quantum gravity.  An active research subject is to identify the possible divergences of spin foam models, or alternatively to show that models are finite. We will discuss in detail the (non--occurrence of) divergences in the Barrett-Crane model, formulated as an integral of delta function weights only. We will furthermore present a simple method to estimate the divergence degree of the so-called bubbles for general spin foam models.

 Divergences in spin foams are expected to be related to the existence of gauge symmetries (diffeomorphisms). Thus we have to conclude that such gauge symmetries are not (fully) present in the model we consider. But we will identify a class of gauge symmetries which occur at special solutions of the equations imposed by the delta function weights. This situation is surprisingly similar to the case of broken diffeomorphism symmetries in discrete gravity, which are present around flat solutions.  We introduce a method to derive (Ward-identity-like) equations for the vertex amplitude of the model in the case of broken gauge symmetries.

\end{abstract}

\medskip

\keywords{spin foam models, bubble divergence, diffeomorphism symmetry}

\maketitle

\section{Introduction}

Spin foam models arose as a path integral approach to quantum gravity \cite{reisenberger}. One of the first specific 4D gravity models constructed was the Barrett--Crane (BC) model \cite{BC}, a more modern alteration of which is \cite{BO}. The EPRL/FK models \cite{EPRL} were also recently proposed due to arguments that the Barrett--Crane model fails certain tests to be a viable model for gravity \cite{gravprop}.

All models underly a unifying construction principle \cite{perezLR}, which in our view is quite accessible in the holonomy representation \cite{holonomy-spinfoam}. There the common starting point is the use of distributions (delta functions) as weights for the faces of the 2--complex on which spin foam models are defined. These delta function face weights are altered by edge functions  convoluted into the face weights. These edge weights are in general also distributional. For the BC model these are again delta functions (for a specific choice of what is termed edge weight or measure factors), and this will allow us to evaluate quite explicitly certain configurations.

Given that the models involve distributional objects one has to worry about divergences and the question arises whether there is need for a regularization. This is an actively studied issue \cite{perez-BC-bubble,simone-spin0,riello} relevant for the definition and behaviour  of group field theories \cite{gft}, for which spin foam models provide the Feynman amplitudes, and the regularization of spin foams in itself.

In this work we will consider in detail the possible divergences that can occur in the Barrett-Crane model (with a specific choice of edge and face weights, which has not been considered before). We will also present a simple method to estimate the occurrences of (single bubble) divergences in general spin foam models (including EPRL/FK). This will show explicitly how the choice of face and edge weight factors influences the divergence properties of the models. Here it will turn out that one has to choose between the invariance of the model under certain edge and face subdivisions and convergence. If one wants to avoid single bubble divergences (divergences related to diffeomorphism symmetry would be expected from multiple bubble configurations) one might also take convergence considerations into account in the determination of these factors \cite{bahr,facef,alexmartin}.

Apart from the possible need for regularization there is another strong motivation to study the divergence structure of spin foams. This is the relation between divergences, the redundancies of delta--distributions and diffeomorphism symmetry \cite{louapre,dittrichreview,cellular-bf, homology-bf,measure-sebastian}, and the proposal that spin foams act as projectors onto the Hamiltonian and diffeomorphism constraints \cite{proj, noui}. As diffeomorphism symmetry leads to non--compact orbits (for vanishing cosmological constants) one would expect that an anomaly free implementation \cite{alexmartin} of these symmetries would lead to divergences due to the integrations of amplitudes over the non--compact orbits (on which these amplitudes are constant). Indeed this relation is well understood for the 3D Ponzano--Regge model \cite{louapre,cellular-bf}, where the diffeomorphism symmetry is implemented as a translation symmetry on the vertices of the triangulation. This symmetry also allows the derivation of recursion relations \cite{recursion} that can be related to the Hamiltonian \cite{wdw3d,recursion15j}.

The issue is more involved for the 4D (gravitational) models. Spin foams can be seen as a discretization of the path integral. Thus the question arises whether diffeomorphism symmetry is preserved under discretization even on the classical level. The single simplex amplitudes of the spin foam models approach the Regge action in the large $j$ limit \cite{asymp}.  The 4D Regge action, as a discretization of the Einstein Hilbert action  does however break in general diffeomorphism symmetry  (as opposed to the 3D Regge action) \cite{broken-sym}.
This statement has however its exceptions \cite{broken-sym,measure-sebastian}: The subdivision of a 4--simplex into five simplices by  placing an inner vertex in the inside of the initial vertex leads to a configuration with vertex translation symmetry. This also leads to  (classical) first class Hamiltonian constraints \cite{broken-sym,hoehn,vbbd} for four--valent vertices in a triangulated three--dimensional hypersurface. Thus one could expect divergences for the spin foam models at least for these configurations. The corresponding symmetry (for instance in the form of redundant delta--functions) could then be used to derive recursion relations and a quantum Hamiltonian.

We will however argue that divergences for the BC model only appear for very special configurations, that first of all have to include two--valent faces (i.\ e.\ faces with only two edges) and furthermore have to combine these two--valent faces in a specific way. (Indeed a gauge symmetry can be found for the case that two two--valent faces are glued onto each other. This is the only configuration for which we found a divergence.) Thus we do not expect a full gauge symmetry related to the subdivision of a simplex, which does not involve two--valent faces.

The subdivision of a simplex corresponds to a situation where  all\footnote{Here we assume that all classical solutions for this case are flat. There could be some special solutions which correspond to discretization artifacts however \cite{broken-sym}.} (classical) solutions are connected by a gauge symmetry. Another case is the occurrence of special (i.e. flat) solutions on more general triangulations. The Hessian around these solutions will feature null modes which signifies the existence of gauge symmetries around these special solutions.

Such symmetries have not been discussed for spin foams so far. Here we will consider an analogue situation for spin foams, that is analyze (gauge) symmetries that occur around special solutions. This is also the reason for considering mostly the BC model in this paper, as (with our choice of edge and face weights) it can be rewritten as the integral over a space of flat connections, i.e. a partition function with only delta function weights. The special solutions are special points in the space of flat connections.
In this work we will present a method to derive recursion relations for the vertex amplitude of the BC model, the $10j$ symbol, which are derived from these special solutions.


~\\

The structure of the paper is as follows. After introducing the Barrett-Crane model with our choice of face and edge weight factors in section \ref{sec:BC}, we will switch to a group integral formulation in section \ref{sec:groupi}. This will introduce effective face weights which capture the possible divergences of spin foam models. For the BC model these effective face weights can be evaluated explicitly and we find them finite for faces with more than two edges (modulo divergences which  occur on measure zero sets). After considering the square of such effective weights, showing that the measure zero set divergences do not matter in this case, we continue with a discussion of so--called bubble divergences in sections \ref{sec:single-bubble}. The methods used there can be generalized to other models as well. As we will show in section \ref{sec:singlebubbleG} this allows for a simple estimate of possible divergences occurring in spin foam models. We then discuss the multiple bubble case, in particular the 4-dipole configuration, important for group field theories, in section \ref{sec:4-dipole}. To this end we will reformulate the partition function as an integral over a space of flat connection. This technique will be essential for the consideration of gauge symmetries, which for the BC model occur around special solutions. We will use this technique in order to discuss these gauge symmetries in section \ref{sec:recurs} and use these symmetries in order to derive recursion relations for the 10j symbols. We close with a discussion and outlook in section \ref{discussion}. The appendix \ref{app:su2calculus} includes some necessary basics on the group $\SU(2)$.

\section{The Barrett-Crane model}\label{sec:BC}

\subsection{Presentation}

In this section we will shortly introduce the model we will be considering in the rest of the paper.

Let $\Gamma$ be a two-complex, $\Gamma_i$ its set of $i$-cells. We call 0-cells vertices, 1-cells lines and 2-cells faces. The spin representation of the Barrett-Crane (BC) model is as follows. A state is an assignment of spins $\{j_f \in \N/2\}_{f\in\Gamma_2}$ to faces. To simplify, assume $\Gamma$ is the 2-skeleton of the dual to a four-dimensional triangulation. Each vertex is dual to a 4-simplex and each line to a tetrahedron. A 4-simplex has five boundary tetrahedra, so that a vertex in $\Gamma$ has degree 5. The faces of $\Gamma$ are dual to triangles. Consider a vertex in $\Gamma$, and denote the incoming lines $a=1,\dotsc,5$. There are ten faces which are identified as the pairs of lines, $(a,b)$ for $1\leq a<b\leq 5$ (and corresponding to the ten triangles of the dual 4-simplex). Each vertex receives a weight, known as the 10j-symbol, labeled by the ten spins of the faces
\be \label{10j}
\{ \textrm{10}j_{ab} \} = \int_{\SU(2)^5} \prod_{a=1}^5 dh_a\ \prod_{1\leq a<b\leq 5} \chi_{j_{ab}}(h_a^{-1}\,h_b) = \begin{array}{c} \includegraphics[scale=0.4]{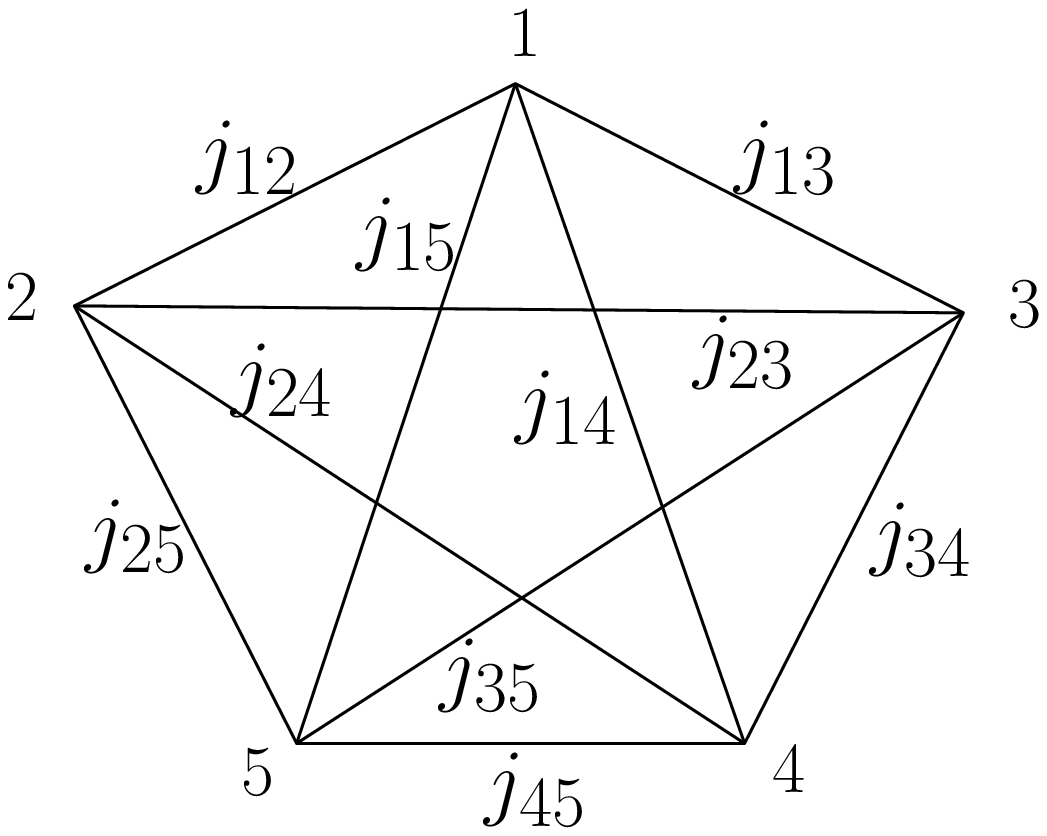} \end{array}.
\ee
Here $\chi_j$ is the $\SU(2)$ character in the representation of spin $j$, $\chi_j ( e^{i\theta\, \hat{n}\cdot \vec{\sigma}}) = \sum_{m=-j}^j e^{im\theta} = \sin (d_j\theta)/\sin\theta$, with $\hat{n}$ a normalized 3-vector, $\vec{\sigma} = (\sigma_x,\sigma_y,\sigma_z)$ the vector formed by the Pauli matrices and the notation $d_j\equiv 2j+1$. The partition function of the model is a state-sum,
\be \label{spin rep}
Z_{\rm BC} = \sum_{\{j_f\}_{f\in\Gamma_2}} \prod_{f\in\Gamma_2} A_f \prod_{e\in\Gamma_1} A_e \prod_{v\in\Gamma_0} \{10j\}.
\ee
Several choices can be found in the literature for the measures on faces $A_f$ and lines $A_e$. We will stick to the following choice,
\be \label{face line measures}
A_f = d_{j_f}^2,\qquad \text{and}\qquad A_e = \frac{1}{\prod_{f\supset e} d_{j_f}}.
\ee

\subsection{Finiteness}

The above choice of measure makes the model quite convergent. This can be found as follows. There are as many sums as faces in $\Gamma$. First we collect the factors $d_{j_f}$ for each sum. The face measure brings $d_j^2$ for any face. For a face with $n$ vertices, dispatching the line measure $A_e$ on the faces yields a factor $d_j^{-n}$. Next we have to deal with the 10j-symbols. Previous studies strongly suggest a bound of the type
\be
|\{10j\}| \leq K\ \prod_f d_{j_f}^{-\alpha},
\ee
for some positive $\alpha$. In \cite{baez-numerics}, numerical evidence gives $\alpha=1/5$. Therefore,
\be \label{boundedBC}
|Z_{\rm BC}| \leq \sum_{\{j_f\}_{f\in\Gamma_2}} \prod_{f\in\Gamma_2} d_{j_f}^{2-n_f} \prod_{v\in\Gamma_0} |\{10j\}| \leq K' \prod_{f\in\Gamma_2} \left[\sum_{j_f} d_{j_f}^{2-n_f -\alpha n_f}\right].
\ee
As a result, if all faces have at least $n_f\geq 3$ vertices, the partition function is finite. However, this result is far from satisfying as it does not provide any insight for quantum gravity. Here are some questions that it leaves unanswered.
\begin{itemize}
\item The above arguments fails in the presence of faces with two vertices only. Such faces appear in generalized triangulations, in particular in the melonic sector which dominates group field theories. We would like to be able to treat them and understand why they are more likely to bring divergences from the quantum gravity point of view.

\item Any spin foam model can be made finite (or arbitrarily divergent) by adding negative (positive) powers of $d_j$ in the face and line measures. Several such different versions of the Barrett-Crane model exist \cite{oriti-boundaryterms, perez-BC-bubble}. We will have to justify our choice \eqref{face line measures}.

\item We would like to formulate the model on arbitrary two-complexes, not only those 2-skeleta dual to regular triangulations. This is necessary to understand bubble divergences (when only part of $\Gamma$ is taken into account).

\item In the BF model, the Wilson loops are constrained to be trivial, and typical divergences come from redundancies in the set of constraints. When turning it to the BC model, are there constraints left? Is the above finiteness result related to absence of constraint redundancies?

\item These redundancies in the BF model are associated to the existence of gauge symmetries which survive on the lattice (and are not gauge-fixed). Does the above finiteness result imply the absence of such gauge symmetries in the BC model?

\end{itemize}
We will answer these questions in the remaining of the paper, through a group integral formulation of the model, where sums over spins are traded for group integrals\footnote{Not surprisingly, integrals are easier to evaluate than sums.}.

\section{Group integral formulation}\label{sec:groupi}

\subsection{Effective face weights}

There exists a group integral representation of this model, which comes out quite naturally as a derivation of the model from a path integral for discretized general relativity \cite{lagrangian-BC}. To each pair line-vertex $(e,v)$ (where $v$ is a vertex of $e$) we associate an $\SU(2)$ element $h_{ev}$ (they equivalently live on half-lines). They have the following geometric meaning. Each 4-simplex of the triangulation (dual to a vertex) is flat and equipped with a local Euclidean frame in $\R^4$. Each boundary tetrahedron, which is dual to a line adjacent to the vertex, spans a 3-dimensional subspace of $\R^4$, determined by the normal to it in the frame of the simplex, denoted $N_{ev}$ (with notations on $\Gamma$). This is a unit vector, hence an element of the 3-sphere. Using the isomorphism between the 3-sphere and $\SU(2)$, the normal $N_{ev}$ is represented\footnote{The map is $N_{ev}^I = \tr (h_{ev}\sigma^I)$, for $I=0,1,2,3$, $\sigma^0 = \mathbbm{I}$, and $\sigma^i$ the Pauli matrices.} as the element $h_{ev}$ of $\SU(2)$.

The interest of such a representation of the normal is that geometric quantities can be expressed through the group law. The dihedral angle between two tetrahedra $e_1,e_2$ in a 4-simplex $v$ is the scalar product between their normals which can be written
\be \label{def dihedral angle}
\cos \theta_{e_1 e_2} = - \tr\, h_{e_1 v}^{-1} h_{e_2 v},
\ee
in the fundamental matrix representation of $\SU(2)$.

A face $f\in \Gamma_2$ can have an arbitrary number $n$ of lines and vertices on its boundary. We divide the face into \emph{wedges}, where a wedge is a pair $(f,v)$ or equivalently a pair of half-lines attached to $v$. There are $n$ wedges $w_{1f},\dotsc,w_{nf}$ around $f$. The product of group elements entering \eqref{def dihedral angle} is canonically associated to a wedge $w$ and to simplify notations we write
\be
h_w = h_{e_1 v}^{-1} h_{e_2 v}.
\ee
The class angle of $h_w$ is therefore the dihedral angle between the tetrahedra dual to the half-lines. We can now re-write the partition function as
\be \label{hol rep}
Z_{\rm BC} = \int \prod_{(e,v)} dh_{ev}\ \prod_{f\in\Gamma_2} \omega(h_{w_{1f}},\dotsc, h_{w_{nf}}),
\ee
The function $\omega$ has $n$ arguments and is called the \emph{effective face weight}. It is fully determined by our choice of measures $A_f,A_e$ to be
\be \label{BCfaceweight}
\omega(h_1,\dotsc,h_n) = \int_{\SU(2)^n} \prod_{i=2}^n d\gamma_i\quad \delta\bigl(h_1\ \gamma_2 h_2 \gamma_2^{-1} \dotsm \gamma_nh_n\gamma_n^{-1}\bigr),
\ee
where $\delta$ is the Dirac delta over $\SU(2)$.

To prove the equivalence between the spin representation \eqref{spin rep} and the group representation \eqref{hol rep} of the partition function, we start by expanding $\omega$ onto the $\SU(2)$ modes, using $\delta(h) = \sum_{j\in \N/2} d_j \chi_j(h)$. A few more formula are needed in order to integrate the elements $\gamma_i$ in \eqref{BCfaceweight}. They are given in the Appendix \ref{app:su2calculus}. The character of products of group elements expands onto the Wigner matrices $D^{(j)}$ as follows $\chi_j(h_1 h_2) = \sum_{m,n=-j}^j D^{(j)}_{mn}(h_1) D^{(j)}_{nm}(h_2)$. Combining this with the orthogonality of the matrix elements of the Wigner matrices \eqref{orthogonality}, we get
\be \label{omega characters}
\omega(h_1,\dotsc,h_n) = \sum_j d_j^{2-n}\ \chi_j(h_1)\,\chi_j(h_2) \dotsb \chi_j(h_n),
\ee
then
\be
Z_{\rm BC} = \sum_{\{j_f\}} \int \prod_{(e,v)} dh_{ev}\ \prod_{f\in\Gamma_2} d_{j_f}^{2-n_f}\ \prod_{w\subset f} \chi_{j_f}(h_w).
\ee
The factor $d_{j_f}^2$ gives the face measure $A_f$, the factors $d_{j_f}^{-n_f}$ are associated to pairs $(e,f)$ and can be re-arranged as the line measure $A_e$. Finally the product of characters over faces can be re-organized as a product over vertices, yielding \eqref{spin rep}.

A first outcome of this formulation is that it directly makes sense on arbitrary two-complexes, not necessarily dual to a regular triangulation.

The effective weight $\omega$ is obviously well-defined as a distribution. Let us integrate $\omega$ with some regular test function,
\be \label{integrate omega}
\begin{aligned}
&\int \prod_{i=1}^n dh_i \int \prod_{i=1}^n d\gamma_i\ \psi(h_1,\dotsc,h_n)\ \delta\bigl(\gamma_1 h_1 \gamma_1^{-1}\ \gamma_2 h_2 \gamma_2^{-1} \dotsm \gamma_nh_n\gamma_n^{-1}\bigr)\\
&=\int \prod_{i=1}^{n-1} dh_i \prod_{i=1}^{n} d\gamma_i\ \psi(h_1,\dotsc,h_{n-1},\gamma_n^{-1}\, (\gamma_{n-1} h_{n-1}^{-1} \gamma_{n-1}^{-1}) \dotsm (\gamma_1 h_1^{-1}\gamma_1^{-1})\, \gamma_n)\\
&= \int \prod_{i=1}^{n-1} dg_i \prod_{i=1}^{n-1} d\gamma_i\ \psi(\gamma_1^{-1} g_1 \gamma_1,\dotsc,\gamma_{n-1}^{-1} g_{n-1} \gamma_{n-1}, \gamma_n^{-1} g_{n-1}^{-1}\dotsm g_1^{-1} \gamma_n).
\end{aligned}
\ee
In the first equality, we have used the Dirac delta to integrate $h_n = \gamma_n^{-1}\, (\gamma_{n-1} h_{n-1}^{-1} \gamma_{n-1}^{-1}) \dotsm (\gamma_1 h_1^{-1}\gamma_1^{-1})\, \gamma_n$. In the last line, we have changed variables to $g_i = \gamma_i h_i \gamma_i^{-1}$, using the translation invariance of the Haar measure (this line simply is a re-writing).

This choice of effective face weight is natural from the way spin foam models are built from BF theory. Indeed, the effective face weight for BF theory is $\omega_{\rm BF} (h_1,\dotsc,h_n) = \delta( h_1\dotsm h_n)$. The insertion of the group elements $\gamma$ is due to the simplicity constraints which break the topological nature of the theory. The way this is implemented in our version of the BC model is interesting because it does not change the functional form of the face weight: it is still formulated with a Dirac delta. A change in the face and line measures \eqref{face line measures} would change the Dirac delta to some other distribution. Moreover, the form of $\omega$ allows to directly draw two conclusions.
\begin{itemize}
\item Just like in the BF model, the potential divergences would come from the fact that multiplication of deltas may not even be defined as a distribution. More precisely, some deltas may be redundant: their arguments are automatically the identity of $\SU(2)$ once the other deltas are satisfied\footnote{In the third reference of \cite{homology-bf}, it was noticed that even in the absence of redundancies, the amplitude may not be finite in BF theory, due to singularities on the set of solutions to the constraints. This issue will be also discussed later in the paper.}. Having Dirac deltas in $\omega$ makes this choice quite convenient to study. In particular, the fact that $\omega_{\rm BF}$ is a delta is the reason why divergences in spin foams are called bubble divergences (because typically there is a redundancy for each `bubble', i.e. independent, spherical, closed surface, so to speak). We will thus be able to compare the contributions of bubbles in this model with the BF case.

\item The expression \eqref{BCfaceweight} further indicates why the model is more likely to be convergent that the BF model, and even why faces with two vertices are the most dangerous with respect to divergences. Since $\SU(2)$ has three real dimensions, the Dirac delta in the effective weight \eqref{BCfaceweight} has three real components. Because the definition of $\omega$ integrates them (over the conjugacy class of each $h_i$), $\omega$ is expected to be more regular, less distributional so to speak, than the $\SU(2)$ delta. The more regular $\omega$ is, the more likely it is that products of $\omega$ are well-defined, removing divergences. Moreover, there is as many integrals in \eqref{BCfaceweight} as vertices around the face. Therefore, $\omega$ certainly becomes quite regular for faces with a sufficient number of lines, and only faces with few lines are expected to be dangerous. We expect this feature to hold more generally in spin foam models for quantum gravity, as the simplicity constraints always amount to smearing the $\SU(2)$ deltas of BF theory, with one integral per line around each face.

\end{itemize}

\subsection{Analysis of the effective face weights}

\subsubsection{Support and geometric interpretation} \label{sec:support}

To understand the support of $\omega$ and the content of the constraint in \eqref{BCfaceweight}, we need a bit of spherical geometry\footnote{The authors are grateful to W. Kaminski for pointing this out.}. A \emph{spherical $n$-gon} is a loop of $n$ geodesic segments on the 2-sphere, with lengths in $[0,\pi]$ (note that it can be degenerate and have self-intersections).

Let $h_1,\dotsc,h_n \in \SU(2)$ with class angles $\theta_k\in[0,\pi]$ defined as $\frac{1}{2}\tr h_k = \cos \theta_k$. Then there exists $\SU(2)$ elements $\gamma_2,\dotsc,\gamma_n$ such that
\be
h_1\ \gamma_2\,h_2\,\gamma_2^{-1}\ \dotsm\ \gamma_n\,h_n\,\gamma_n^{-1} = \mathbbm{I},
\ee
if and only if there exists a spherical $n$-gon with lengths $(\theta_1,\dotsc,\theta_n)$. The fact that the constraint implies the existence of a spherical polygon is proved in \cite{BS} and we will not repeat it. We will prove the reverse. The case $n=2$ is trivial. We proceed by induction on $n$ starting with $n=3$.

Write $h_k = \exp i\theta_k\, \vec{\sigma}\cdot \hat{n}_k$ with $\theta_k\neq 0,\pi$, and consider a spherical triangle with lengths $(\theta_1, \theta_2, \theta_3)$. The angles between the sides of the triangles are given by the spherical law of cosines,
\be \label{spherical-cosines}
\cos\phi_{ij} = \frac{\cos\theta_k - \cos\theta_i\,\cos\theta_j}{\sin\theta_i\ \sin\theta_j}.
\ee
Set $\gamma_2\in\SU(2)$ such that $R(\gamma_2) \hat{n}_2 = \hat{u}_2$ with $\hat{u}_2$ being any unit vector satisfying
\be
\hat{n}_1 \cdot \hat{u}_2 = -\cos \phi_{12}.
\ee
The $\SU(2)$ element $\gamma_2 h_2 \gamma_2^{-1}$ has the same class angle as $h_2$, but its axis is $\hat{u}_2$, i.e. $\gamma_2 h_2 \gamma_2^{-1} = \exp i\theta_2\,\vec{\sigma}\cdot \hat{u}_2$. Then the equation \eqref{spherical-cosines} for $i=1, j=2, k=3$ exactly reads
\be
\tr\, h_1\,\gamma_2 h_2 \gamma_2^{-1} = \tr\, h_3^{-1},
\ee
which implies that the matrices on both sides are conjugated to each other, by, say, $\gamma_3\in\SU(2)$. In other words, there exist $\gamma_2, \gamma_3\in\SU(2)$ such that $h_1 \gamma_2 h_2 \gamma_2^{-1} = \gamma_3 h_3^{-1} \gamma_3^{-1}$.

Completing the induction is easy. Assume there is a $n$-gon with spherical lengths $(\theta_1,\dotsc,\theta_n)$. By splitting it on its $(n-2)$-th vertex, we get a $(n-1)$-gon with lengths $(\theta_1,\dotsc,\theta_{n-2}, \theta_0)$ and a spherical triangle with lengths $(\theta_0, \theta_{n-1}, \theta_n)$. The induction hypothesis ensures the existence of $\SU(2)$ elements such that $h_1 \gamma_2 h_2 \gamma_2^{-1} \dotsm \gamma_{n-2}h_{n-2}\gamma_{n-2}^{-1} h_0 = \mathbbm{I}$ where $h_0$ has class angle $\theta_0$. As for the triangle the above proof of the case $n=3$ shows that there exist $\SU(2)$ elements such that $\tilde{h}_0 \tilde{\gamma}_{n-1} h_{n-1}\tilde{\gamma}_{n-1}^{-1} \tilde{\gamma}_n h_n \tilde{\gamma}_n^{-1}=\mathbbm{I}$, where $\tilde{h}_0$ has also class angle $\theta_0$. Therefore $h_0$ and $\tilde{h}_0$ are conjugated and we can write for the triangle $h_0^{-1} \gamma_{n-1} h_{n-1} \gamma_{n-1}^{-1} \gamma_n h_n \gamma_n^{-1} = \mathbbm{I}$. This completes the proof.
~\\

Furthermore, existence of an $n$-gon is equivalent to the spherical polygon inequalities on $(\theta_1,\dotsc,\theta_n)$. This means that those inequalities provide the support of the effective face weight. They read
\be
\sum_{i\in P} \theta_i- \sum_{i\in P'} \theta_i - \pi (|P|-1) \leq 0,
\ee
for any subset $P\subseteq \{1,\dotsc,n\}$ with $|P|$ odd, and $P'=\{1,\dotsc,n\} \setminus P$.

In summary the effective face weights are supported on configurations of group elements $h_1,\cdots,h_n$ for which the corresponding class angles define a (possibly degenerate) spherical $n$--gon.  We will find this constraint again in the explicit expression of the face weights.

\subsubsection{Explicit expression of the face weight}

From the bound \eqref{boundedBC}, we see that only faces with two vertices are dangerous, suggesting that the effective face weight of faces with at least three vertices are quite regular.

\paragraph{Face with two vertices, $n=2$.} In that case, $\omega(h_1,h_2)$ is actually well-known to be the one-dimensional delta constraining $h_1$ and $h_2$ to lie in the same conjugacy class (i.e. to have the same rotation angle),
\be
\omega(h_1,h_2) = \int d\gamma \ \delta(h_1\,\gamma\,h_2\,\gamma^{-1}) = \sum_j \chi_j(h_1)\,\chi_j(h_2) = \frac{1}{\sin\theta_1\,\sin\theta_2} \sum_{k\in\Z} \sin k\theta_1\,\sin k\theta_2.
\ee

\paragraph{Face with more than two vertices, $n\geq 3$.} Since $n=2$ has only a one-dimensional delta function, and there are more integrals over conjugacy classes for $n\geq 3$, $\omega$ does not contain any delta anymore and is regular almost everywhere. To make this precise, the expression of the character in terms of the class angle is inserted into \eqref{omega characters},
\be\label{19}
\begin{aligned}
\omega(h_1,\dotsc,h_n) &= \sum_{k=1}^\infty \frac1{k^{n-2}}\,\frac{\sin (k\theta_1) \dotsb \sin(k\theta_n)}{\sin\theta_1\dotsb \sin\theta_n},\\
&= \frac{1}{(2i)^n} \sum_{k=1}^\infty \frac1{k^{n-2}} \sum_{\epsilon_2,\dotsc,\epsilon_n=\pm1} \frac{\epsilon_2\dotsb \epsilon_n}{\sin\theta_1\dotsb \sin\theta_n}\,\Bigl[ e^{ik(\theta_1+\sum_{l=2}^n\epsilon_l\theta_l)} + (-1)^n\,e^{-ik(\theta_1+\sum_{l=2}^n\epsilon_l\theta_l)} \Bigr].
\end{aligned}
\ee
If one group element is set to $h_i=\pm\mathbbm{I}$ (i.e. $\theta_i=0,\pi$), then $\omega$ reduces to the face weight with simply one argument less, $n\rightarrow n-1$ (which is the weight for a face with one line and one vertex less). Therefore we assume that all $h_i\neq \pm \mathbbm{I}$.

This allows to evaluate the sum over $k$ for all terms independently,
\be\label{20}
\frac1{(2i)^n} \sum_{k=1}^\infty \frac1{k^{n-2}} \Bigl[ e^{ik\Theta} + (-1)^n\,e^{-ik\Theta} \Bigr] = \frac1{2^n} \sum_{k=1}^\infty \frac1{k^{n-2}} \,\cos \Bigl( k\Theta -n\frac{\pi}{2}\Bigr) = \frac{\pi^{n-2}}{8\,(n-2)!}\ B_{n-2}\Bigl(\frac{\Theta}{2\pi}\Bigr),
\ee
where we have recognized\footnote{We could also recognize the real and imaginary parts of the polylogarithm $\Li_n(z) = \sum_{k=1}^\infty z^k/k^s$.} the Fourier expansion of the Bernoulli polynomial $B_{n-2}$, which holds for $\Theta\in[0,2\pi]$.  If $\Theta$ is outside this interval we need to shift it back to be in $[0,2\pi]$ using that the left hand side has to be periodic in $\Theta$.

Notice that (\ref{20})  is absolutely convergent for $n\geq 4$. For $n=3$ we get $\sum_{k=1}^\infty \sin(k\Theta)/k$ which gives the Fourier expansion of the sawtooth wave, i.e. a finite function which is not continous. This applies to each configuration of $\epsilon_k$, appearing in (\ref{19}), with $\Theta_{\{\epsilon\}} = \theta_1+\sum_{l=2}^n\epsilon_l\theta_l \mod(2\pi)$. Thus we have
\be
\omega(h_1,\dotsc,h_n) = \frac{\pi^{n-2}}{8\,(n-2)!} \frac1{\sin\theta_1\dotsb \sin\theta_n} \sum_{\epsilon_2,\dotsc,\epsilon_n=\pm1} \epsilon_2\dotsb \epsilon_n\ B_{n-2}\Bigl(\frac{\Theta_{\{\epsilon\}}}{2\pi}\Bigr).
\ee


The Bernoulli polynomial $B_n(x)$ has a monomial of highest order $x^n$. However, the sum over the signs $\epsilon_k=\pm$ may lead to simplifications.

Let us focus on the case $n=3$. There, we need the polynomial $B_1(\frac{\theta+2\pi N}{2\pi}) = \frac{\theta}{2\pi} +N -1/2$, where $N\in \mathbb{Z}$ has to be chosen such that $\theta+2\pi N\in [0,2\pi]$. Assume that $\theta_1,\theta_2,\theta_3$ satisfy the spherical triangle inequalities, i.e. $\theta_1+\theta_2+\theta_3 \leq 2\pi$ and $\theta_a\leq \theta_b+\theta_c$ for any permutation of $a,b,c=1,2,3$. Of the four combinations of $\epsilon_2,\epsilon_3$, three arguments $\theta_1+\sum_{l=2}^n \epsilon_l \theta_l$ are in $[0,2\pi]$. However for $\epsilon_2=\epsilon_3=-1$ we obtain a negative argument. For this last summand we need to shift the argument back to $[0,2\pi]$ by choosing $N=1$. Then in the sum over the $\epsilon_2,\epsilon_3$ all the linear terms in $\theta_k$ vanish and we are left with a constant
\be
B_1\Bigl(\frac{\theta_1+\theta_2+\theta_3}{2\pi}\Bigr) - B_1\Bigl(\frac{\theta_1+\theta_2-\theta_3}{2\pi}\Bigr) - B_1\Bigl(\frac{\theta_1-\theta_2+\theta_3}{2\pi}\Bigr) + B_1\Bigl(\frac{\theta_1-\theta_2-\theta_3+2\pi}{2\pi}\Bigr) = 1 \quad .
\ee
Hence if the spherical triangle inequalities are satisfied, the effective face weight reduces to
\be
\omega(h_1,h_2,h_3) = \frac{\pi}{8}\ \frac1{\sin\theta_1\ \sin\theta_2\ \sin\theta_2}.
\ee
Consistently with the support found in the Section \ref{sec:support}, one can check that violations of the triangle inequalities lead to $\omega =0$. Thus the face weight for a three--valent face is simply
\be\label{24}
\omega(h_1,h_2,h_3) = \frac{\pi}{8}\ \frac1{\sin\theta_1\ \sin\theta_2\ \sin\theta_2}  H(\theta_1,\theta_2,\theta_3)
\ee
where $H(\theta_1,\theta_2,\theta_3)=1$ if the spherical triangle inequalities are satisfied and vanishing otherwise.

For higher--valent faces, with $n$ edges, the Heaviside like function $H$ in (\ref{24}) is replaced by some piecewise  polynomial of order $(n-3)$  so that the support is on configurations satisfying the spherical triangle inequalities.

As a conclusion, the face weight is a function of the class angles only. The face with two vertices is the most singular, and the weight becomes smoother as the number of vertices per face increases. This is expected because the number of group averages in the expression \eqref{BCfaceweight} for $\omega$ is precisely the number of vertices. As the group averaging comes from the simplicity constraints, we expect this feature to also hold for the EPR/FK model. We will provide evidence (which will depend on the choice of certain edge weight factors) in section \ref{sec:singlebubbleG}.

Let us note that possible divergences of the models can be also analyzed with the help of microlocal analysis \cite{fw}. Wave front sets are a refinement (or extensions to co--tangent space) of the singular support of a function. The wave front sets for the effective face weights for the BC model are non--empty and indeed reflect the configurations at which the effective face weights are non-smooth \cite{private}. Thus  wave front sets in spin foam models do not necessarily correspond to divergences but could also just signify non--smooth behaviour. In this case there is no need for regulating the models.


\subsection{Two faces glued together: the square of the face weight} \label{sec:twofaces}

Although pretty explicit, the effective face weights are functions  of the dihedral angles, which involve in their definition the product of two group elements: $ \cos \theta_{e_1 e_2} = - \tr\, h_{e_1 v}^{-1} h_{e_2 v}$. Thus if we glue faces together we would have to disentangle these group elements again, as the $h_{e_1 v}$ and $h_{e_2 v}$ will in general be shared by different sets of faces.

Also, even for effective face weights with valency larger than two, the sets where some $h_i$ are $\pm\mathbbm{I}$ might lead to (delta--function like) singularities (this is what prevents $\omega$ from being a standard function) 
and they may contribute, though having zero measure. Therefore, for practical calculations, we will rather use the expression \eqref{BCfaceweight} and deal with products of $\SU(2)$ delta functions.


We have already emphasized that typical divergences might show up because of the products of deltas are a priori ill-defined, due to redundancies of the enforced constraints. Such redundancies typically are expected when some faces are glued in a way that creates closed surfaces in $\Gamma$ (boundary of 3-cells if $\Gamma$ is the 2-skeleton of a higher dimensional cell complex). In the BF spin foam model, it is easy to check that the presence of 3-cells is associated to a redundant Dirac delta, whose argument is automatically the unit of the group whenever the deltas on the other faces are satisfied\footnote{The full evaluation of divergences in the BF spin foam models requires in addition to take into account the reducibility of the gauge symmetries and global, topological effects, see \cite{bubblediv3}.}. In the following subsections, we will focus on these typical situations and argue that there is no divergence, expect in one case where two faces with exactly two vertices are glued together.

The first step to study this type of situation is to look at the square of $\omega$. Geometrically, that corresponds to a closed surface in $\Gamma$ made of two faces with the same boundary lines and vertices. In the BF case, $\omega=\delta$ and its square is obviously \emph{not} well-defined, because the two deltas impose the same constraint twice. Here the question is therefore: is $\omega$ a distribution which can be squared?

Consider that there are $n$ lines around the two faces. There are also other faces which may share boundary lines with the two faces. Therefore the amplitude on $\Gamma$ reads
\be
Z(\Gamma) = \int \prod_{a=1}^n dh_a\ \bigl[\omega(h_1,\dotsc,h_n)\bigr]^2\ f(h_1,\dotsc,h_n),
\ee
where $f$ is the result of integrating all the group elements $h_{ef}$ in $\Gamma$ at fixed wedge group elements $h_1,\dotsc,h_n$ on the boundary of the two faces. $f$ is typically a distribution, or it might itself contain divergences. However, our aim is to isolate the contribution of the two faces glued together and therefore we consider $f$ as a regular function. The integral is on a compact manifold which implies
\be
Z(\Gamma) \leq K \int \prod_{a=1}^n dh_a\ \bigl[\omega(h_1,\dotsc,h_n)\bigr]^2,
\ee
where the constant $K$ is the maximal value of $f$ on $\SU(2)^n$.

It is possible to calculate the integral of $\omega^2$ exactly. Going through the steps of the Equation \eqref{integrate omega},
\be
\int \prod_{a=1}^n dh_a\ \bigl[\omega(h_1,\dotsc,h_n)\bigr]^2 = \int  \prod_{a=1}^n d\beta_a\,d\gamma_a\ \prod_{b=1}^{n-1} dg_b\ \delta\Bigl( (\beta_1\gamma_1^{-1} g_1 \gamma_1 \beta_1^{-1})\ (\beta_2 \gamma_2^{-1} g_2 \gamma_2 \beta_2^{-1}) \dotsb \beta_n \gamma_n^{-1} g_{n-1}^{-1} \dotsb g_1^{-1} \gamma_n \beta_n^{-1} \Bigr)
\ee
Re-absorbing $\gamma_i^{-1}$ on the right of $\beta_i$ gives
\be \label{Ztwofaces}
\int \prod_{a=1}^n dh_a\ \bigl[\omega(h_1,\dotsc,h_n)\bigr]^2 = \int  \prod_{a=1}^n d\beta_a \prod_{b=1}^{n-1} dg_b\ \delta\Bigl( (\beta_1 g_1 \beta_1^{-1})\ (\beta_2 g_2 \beta_2^{-1}) \dotsb (\beta_{n-1} g_{n-1} \beta_{n-1}^{-1})\ \beta_n g_{n-1}^{-1} \dotsb g_1^{-1} \beta_n^{-1} \Bigr)
\ee
We then use the character expansion and explicit integration of Wigner matrices. Doing so yields
\be
\begin{aligned}
\int \prod_{a=1}^n dh_a\ \bigl[\omega(h_1,\dotsc,h_n)\bigr]^2 &= \sum_{j\in\N/2} \frac1{d_j^{n-2}} \int \prod_{a=1}^{n-1} dg_a\ \chi_j(g_1\dotsm g_{n-1}) \prod_{a=1}^{n-1} \chi_j(g_a),\\
&= \sum_{j\in\N/2} \frac1{d_j^{2(n-2)}} = \zeta(2n-4).
\end{aligned}
\ee
This is obviously finite as soon as $n\geq3$. One concludes that $Z(\Gamma)$ is finite for regular enough $f$.

{\bf Remark.} The above result is identical to the partition function of 2d BF, i.e. 2d Yang-Mills at zero coupling, on a surface of genus $n-1$. This is no coincidence since \eqref{Ztwofaces} can actually be re-written in the typical 2d YM form. We first abosrb $\beta_n$ into the other $\beta_a \leftarrow \beta_n^{-1} \beta_a$. Then we proceed to the change of variables $(g_a,\beta_a) \mapsto (k_a,\alpha_a)$ step by step starting with $a=1$,
\begin{alignat*}{3}
g_1 &= g_{2} \dotsm g_{n-1}\ k_1^{-1}\ g_{n-1}^{-1} \dotsm g_{2}^{-1}& \quad&\text{and}\quad &\beta_1 &= g_{2} \dotsm g_{n-1}\ \alpha_1\ g_{n-1}^{-1} \dotsm g_{2}^{-1},\\
g_2 &= g_{3} \dotsm g_{n-1}\ k_2^{-1}\ g_{n-1}^{-1} \dotsm g_{3}^{-1}& \quad&\text{and}\quad &\beta_2 &= g_{3} \dotsm g_{n-1}\ \alpha_2\ g_{n-1}^{-1} \dotsm g_{3}^{-1},\\
 &\mathrel{\makebox[\widthof{=}]{\vdots}}& \quad&\text{}\quad& &\mathrel{\makebox[\widthof{=}]{\vdots}} \\
g_{n-2} &= g_{n-1}\ k_{n-2}^{-1}\ g_{n-1}^{-1}& \quad&\text{and}\qquad &\beta_{n-2} &= g_{n-1}\ \alpha_{n-2}\ g_{n-1}^{-1},\\
g_{n-1} &= k_{n-1}^{-1}& \quad&\text{and}\qquad &\beta_{n-1} &= \alpha_{n-1}.
\end{alignat*}
This recasts \eqref{Ztwofaces} in the form
\be
\int \prod_{a=1}^n dh_a\ \bigl[\omega(h_1,\dotsc,h_n)\bigr]^2 = \int \prod_{a=1}^{n-1} d\alpha_a\,dk_a\quad \delta\bigl( [k_1,\alpha_1]\,[k_2,\alpha_2]\,\dotsm\,[k_{n-1},\alpha_{n-1}]\bigr),
\ee
where $[k,\alpha]=k\alpha k^{-1} \alpha^{-1}$ is the group commutator. The argument of the delta function in the above equation is recognized as a $\SU(2)$ version of the presentation of the fundamental group of the surface of genus $n-1$, as expected, \cite{witten-2dym,forman, goldman}.

\section{Finiteness of single bubble contributions} \label{sec:single-bubble}

The calculation performed in the case of two faces glued along their boundaries generalizes to any single bubble of arbitrary shape. The same way we had found the partition function of 2d BF on a surface of genus the number of boundary lines minus 1, we will see that the calculation goes through in the case of a single bubble thanks to the well-known fact that lattice gauge theories are trivial in two dimensions.

\subsection{Single bubble contribution as two-dimensional spin foams} \label{sec:mainassumption}

Suppose we can identify in $\Gamma$ a closed surface $\Sigma$ of arbitrary Euler characteristic $\chi=V-E+F$. It is usually referred to in the quantum gravity literature as a (not necessarily spherical) bubble. It is such that each line is shared by exactly two faces. The partition function on $\Gamma$ reads
\be
Z(\Gamma) = \int \prod_{\substack{(e,v)\\ e,v \subset \Sigma}} dh_{ev}\ \prod_{f\subset \Sigma} \omega(h_{w_{1f}},\dotsc,h_{w_{nf}}) \prod_{\substack{(e,v) \\ e\not\subset \Sigma}} dh_{ev}\ \prod_{f\not\subset \Sigma} \omega(h_{w_{1f}},\dotsc,h_{w_{nf}}).
\ee
Integrating all the group elements $h_{ev}$ where $e$ does not belong to $\Sigma$ produces a function depending on the group elements on $\Sigma$,
\be
f_\Sigma(\{h_{ev}\}_{e,v \subset \Sigma}) = \int \prod_{\substack{(e,v) \\ e\not\subset \Sigma}} dh_{ev}\ \prod_{f\not\subset \Sigma} \omega(h_{w_{1f}},\dotsc,h_{w_{nf}}).
\ee
It is a sort of Hartle-Hawking wave-function on $\Sigma$ but it is not clearly well-defined {\it a priori}. However, to isolate the contribution of $\Sigma$, we will consider that $f_\Sigma$ is a regular function. Because it is defined on a compact space, we assume it is bounded by some constant $K$. Therefore, $Z(\Gamma)$ is bounded by $K$ times the partition function where $f$ is set to 1. This is exactly the partition function of the Barrett-Crane model on the two-dimensional surface $\Sigma$. Therefore
\be
Z(\Gamma) \leq K\ Z(\Sigma),
\ee
with
\be
Z(\Sigma) = \int \prod_{\substack{(e,v)\\ e,v \subset \Sigma}} dh_{ev}\ \prod_{f\subset \Sigma} \omega(h_{w_{1f}},\dotsc,h_{w_{nf}}).
\ee

To evaluate the potentially divergent contribution of $\Sigma\subset \Gamma$, we will calculate $Z(\Sigma)$ instead of $Z(\Gamma)$. This amounts to ignoring the faces external to $\Sigma$. In the usual spin foam language, this is simply setting the spins of the external faces to zero. This has already been used in the literature to isolate divergent contributions. In \cite{simone-spin0}, it was argued to yield a fair evaluation of the divergences, just like in ordinary quantum field theory the loop divergences are often evaluated by setting the momenta of the external legs to zero.

However, it seems to have gone un-noticed that when doing so on a single bubble, the remaining part $Z(\Sigma)$ is just a two-dimensional version of the initial model. Thinking of spin foam models as generalized lattice gauge theories, and given that two-dimensional lattice gauge theories are solvable, this gives us hope to calculate $Z(\Sigma)$ exactly. This is what we do now, in the Barrett-Crane case in the Section \ref{sec:BC2d}, and for generic spin foam models in \ref{sec:singlebubbleG} (the BC case is just a particular case, but we treat them separately because the BC model is our working example in this paper).

\subsection{The Barrett-Crane model on two-dimensional surfaces} \label{sec:BC2d}

To compute $Z(\Sigma)$, we use the character expansion of $\omega$, \eqref{omega characters}, to get
\be
Z(\Sigma) = \sum_{\{ j_f\}_{f\subset \Sigma}} \int \prod_{\substack{(e,v)\\ e,v \subset \Sigma}} dh_{ev}\ \prod_{f\subset \Sigma} d_{j_f}^{2-n_f} \chi_{j_f}(h_{w_{1f}})\dotsb \chi_{j_f}(h_{w_{nf}}),
\ee
where the sum is over all possible assignments of spins to faces, $n_f$ is the number of lines on the boundary of $f$, and $h_w$ is the wedge holonomy. Remember that a wedge is identified by a pair `vertex-face', or equivalently the two lines along that face which meet at that vertex. We denote these two lines $e_w, e'_w$. To perform the integrals explicitly, we notice that they actually factorize onto vertices,
\be
Z(\Sigma) = \sum_{\{j_f\}} \prod_f d_{j_f}^{2-n_f} \prod_v \left[\int \prod_{e \supset v} dh_{ev}\ \prod_{w\supset v} \chi_{j_f}( h_{e_w v}^{-1} \,h_{e'_w v})\right].
\ee
Around each vertex with $n_v$ lines, there are also $n_v$ faces (or rather wedges), and since each line is shared by exactly two wedges, we can label the lines and the wedges, say in clockwise order. At each vertex we have
\be
\int dh_1\dotsm dh_{n_v}\ \chi_{j_{f_1}}( h_1^{-1} h_2)\ \chi_{j_{f_2}}( h_2^{-1} h_3) \dotsm \chi_{j_{f_{n_v}}}( h_{n_v}^{-1} h_1) = \left[\prod_{f,f'} \delta_{j_f,j_{f'}}\right]\ d_j^{2-n_v}.
\ee
We have used the formula \eqref{convolution} $n_v-1$ times to integrate products of characters. As a result, all spins around $v$ are must have the same value which we have denoted $j$. Since $\Sigma$ is connected, the spins of all faces must be identical, so that the sum over all spin assignments reduces to a single sum. The summand is obtained by gathering all powers of $d_j$,
\be
Z(\Sigma) = \sum_{j\in\N/2} \prod_f d_j^{2-n_f} \prod_v d_j^{2-n_v} = \sum_{j\in\N/2} d_j^{2F+2V-\sum_f n_f - \sum_v n_v}.
\ee
On the triangulation of a surface, $\sum_f n_f = \sum_v n_v =2E$, hence
\be\label{38}
Z(\Sigma) = \sum_{j\in\N/2} d_j^{2\chi -2E}.
\ee

{\bf Remark 1 -- Finiteness.} The result is divergent if $E\leq \chi$. Since $\chi\leq 2$ and we want at least $E\geq 2$, we find that the only divergent case is $E=\chi=2$, meaning a spherical bubble with only two lines. This is the case of two faces with two boundary lines glued together already seen in the section \ref{sec:twofaces}. Any other bubble is finite,
\be \label{2dBC}
Z(\Sigma) = \zeta(2E-2\chi) = (-1)^{E-\chi+1}\,\frac{B_{2(E-\chi)}\,(2\pi)^{2(E-\chi)}}{2\,(2E-2\chi)!}.
\ee
Here $\zeta$ is the Riemann zeta function and $B_{n}$ a Bernoulli number.

{\bf Remark 2 -- Invariance.} In the BF case, the partition function on a surface of Euler characteristic $\chi$ is $\zeta(-\chi)$ and is therefore independent of the triangulation. Here $Z(\Sigma)$ depends on the triangulation only through the number of lines and not its particular shape. It means that $Z(\Sigma)$ is invariant under homeomorphisms of the triangulation which preserves the number of lines. In particular, the 2-2 Pachner move does so and therefore leaves $Z(\Sigma)$ invariant.

{\bf Remark 3 -- 4-2 Pachner move.} The section \ref{sec:twofaces} is obviously a particular case of this section, with $\chi=2$ and $E=n$ the number of boundary lines of the two faces. A more interesting case for quantum gravity is the 4-2 Pachner move. As a Pachner move, it is a change of triangulation which preserves the topology of $\Gamma$. In the dual of $\Gamma$, one changes a configuration of two 4-simplices which share a common tetrahedron and thus have eight boundary tetrahedra, with four 4-simplices each contributing to two boundary tetrahedra. The four 4-simplices are glued together in a specific pattern such that two 4-simplices share exactly one tetrahedron. In the 2-complex $\Gamma$, we find 4 vertices completely connected by 6 lines which form 4 triangular faces. This pattern corresponds to the boundary of a tetrahedron and this is precisely the surface $\Sigma$, with $\chi=2, E=6$. Thus the contribution to the partition function from the bubble in the 4-2 Pachner move configuration is finite.

\subsection{Single bubble contributions for general models}\label{sec:singlebubbleG}

Spin foam models in 2D reduce to 2D (standard) lattice gauge theories \cite{holonomy-spinfoam}, implying that the above calculation for a single bubble can be extended to generic models. A wide class of models (including BC with generic choices of edge and face weights and the EPRL/FK model) \cite{holonomy-spinfoam} is given by
\be\label{Zben}
Z=\int_G \prod_{(e,f)} dh_{ef} \, \prod_e C(\{h_{ef}\}_{f\supset e}) \prod_f w(h_{e_1 f} \cdots h_{e_n f}) \quad .
\ee
Here we integrate  over group $G$ elements associated to edge-face pairs $(ef)$. For every edge we have an edge weight $C$ which depends on the group elements $h_{ef}$, for which $e$ is an edge in the boundary of $f$. For each face we have a (`bare') face weight $w$ , which is a class function and evaluated on the holonomy around the face.

For a 2D surface, there are always two faces adjacent to a given edge. The edge weights $C$ have also to satisfy a certain invariance property, which means that for the 2D case we can expand $C$ as follows into irreducible unitary representations $\rho$ of $G$
\be\label{exp1}
C(h_1,h_2)=\sum_\rho \tilde C_{\rho} \dim(\rho)\,\, \chi_\rho(h_1h_2^{-1}) \quad .
\ee
The face weights are expanded as
\be\label{exp2}
w(h)=\sum_\rho \tilde \omega_\rho \dim(\rho) \, \chi_\rho(h)  \quad .
\ee

Using these expansions in (\ref{Zben}) for the 2D case one notices that the sums over the representation labels $\rho$ reduce to one sum as the group integrations impose Kronecker deltas between representation labels. Taking care of all the dimension factors that come from the expansions (\ref{exp1},\ref{exp2}), the group inner product between representation matrix elements,  as well as from the contractions of Kronecker deltas (which gives traces around vertices) we obtain
\be\label{2dcase}
Z_{2D}\,=\, \sum_\rho (\tilde C_\rho)^E (\tilde \omega_\rho)^F (\dim \rho)^{V-E+F} \,\,=\,\, \sum_\rho (\tilde C_\rho)^E (\tilde \omega_\rho)^F (\dim \rho)^\chi \quad .
\ee

For (standard) lattice gauge theory we have $\tilde C_\rho \equiv 1$, and we recover the corresponding 2D partition function. For the BC model we have $G=SU(2)\times SU(2)$  and with our choice of edge and face weights
\be
\tilde \omega_\rho=1 \quad ,\quad\quad \tilde C_\rho= \delta_{\rho, (j,j)} \left(\frac{1}{d_j}\right)^2
\ee
so that we recover (\ref{38}). However we see that changing edge weights, for instance introducing a factor $d_j^2$ per edge, so that $\tilde C_\rho= \delta_{\rho, (j,j)}$  would lead to a triangulation invariant but divergent result for a spherical bubble.


For the EPRL model ($G=SU(2)\times SU(2)$) with Barbero--Immirzi parameter $\gamma$, leaving the face weights $\tilde \omega_\rho$ free for the moment, we have for the edges \cite{holonomy-spinfoam}
\be\label{45}
\tilde C_\rho= \sum_j \frac{\dim(j)}{ \dim(\frac{1+\gamma}{2} j) \dim(\frac{|1-\gamma|}{2} j) } \,\, \delta(\rho, (\frac{1+\gamma}{2} j,  \frac{|1-\gamma|}{2} j  )) \,\,  (d_e(\rho))^2 \quad .
\ee
Here $d_e(\rho)$ is an edge weight factor which is left undetermined. 
We obtain a non--vanishing coefficient only if the  $SU(2)\times SU(2)$ representation $\rho=(j',j'')$ is of the form $(\frac{1+\gamma}{2} j,  \frac{|1-\gamma|}{2} j  )$.


We have two free functions, the face weights $\tilde \omega_\rho$ and the edge weight factors $d_e(\rho)$. The choice of these factors will heavily influence the convergence properties of the model. Different requirements have been proposed to fix these weights \cite{alexmartin,bahr, facef}. With the simple arguments put forward here we can comment on how these requirements will influence the convergence of single spherical bubbles. \\

(i) We can require that the model is invariant under edge subdivisions (here for edges which are shared by only two faces) and face subdivisions. To achieve this we choose $\tilde \omega_\rho=1$ and $d_e(\rho)$ such that $\tilde C_\rho$ equal to one or is vanishing (if $\rho$ is not admissible). This will give a  triangulation invariant 2D model. In this case spherical bubbles will diverge (assuming that there are infinitely many admissible representations). \\

(ii) A weaker requirement is invariance under subdivision of faces.\footnote{Here we mean that a face is subdivided by a two--valent edge which goes between two already existing vertices, i.e. we do not create new vertices. We should point out that there exist other notions of face subdivisions which create new vertices and hence a larger number of additional edges.} Dividing one face into two we raise the number of faces and the number of edges by one. Thus invariance requires $\tilde \omega_\rho = (\tilde C_\rho)^{-1}$ for admissible representations $\rho$. The convergence then depends on the difference between the number of faces and edges. If we consider the square of an effective face weight, it forms a spherical bubble with two faces. As long as $\tilde C_\rho$ scales with some negative power of $\dim \rho$ we obtain a more convergent result with growing number of edges, where the specifics again depend on the edge weight factor.

If $\tilde \omega_\rho = (\tilde C_\rho)^{-1}$ we have a divergent partition functions for all spheres where the number of edges and faces are equal to each other. This includes the bubbles that appear in the 4-dipole configurations. \\

(iii) One can also adjust the face weight and edge weight factors to obtain convergent results for specific families of bubbles.\\

Thus we see that we can get an estimate (as we ignore the contribution of faces which connect to but are not part of the bubble) on the behaviour of bubbles and the behaviour of effective face weights by quite simple methods.  This allows to choose the edge weight factors and face weights according to the divergent or convergent behaviour one wants to achieve.  We have seen however that requiring triangulation independence in 2D  (i.e. invariance under face and edge subdivisions) comes at the cost of divergent spherical bubbles and distributional effective face weights for arbitrary number of edges.

\section{Multiple bubbles: the 4-dipole} \label{sec:4-dipole}

Next we will discuss a configuration with multiple bubbles, known as the 4-dipole. It in particular it arises in discussions of group field theories as configurations which include such dipoles are the most divergent ones \cite{gft2}.  For the BC model the 4-dipole could be divergent due to the appearance of two--valent faces, which glue to (multiple) spherical bubbles. These bubbles have three faces and three edges -- thus the single bubble contribution as discussed in section \ref{sec:mainassumption}, converge. The assumption of this section was to ignore faces external to the bubble. Thus this is also a test whether this assumption holds in this case. Indeed, we will find that divergences do not occur -- at least no divergences due to redundancies of delta functions. (There are however singularities on a measure zero set and we leave the integrability of these singularities open.)

An expansion in spin variables for this case is not sufficient to determine convergence. We will therefore switch to the group representation and basically show that no redundancies arise if we solve for all the delta--functions appearing in the partition function for this configuration (the method is detailed in \cite{homology-bf}).

Let us describe the 4-dipole configuration. We consider a piece of triangulation with two 4-simplices which are glued together along four of their five boundary tetrahedra. These four internal tetrahedra have six triangles and four edges in total, which are thus shared by the two 4-simplices. The boundary of the gluing consists of two tetrahedra. Notice that their triangles are shared with internal tetrahedra (in a 4-simplex, a triangle is shared by exactly two tetrahedra). Therefore, each triangle of a boundary tetrahedron is identified with a triangle of the other boundary tetrahedron. This means that the two 4-simplices actually share all their triangles.

\begin{figure}
\includegraphics[scale=.7]{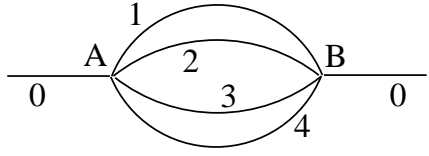}
\caption{\label{fig:4-dipole} The vertices $A, B$ are dual to 4-simplices, which are glued together by four tetrahedra here represented as the lines 1,2,3,4. The lines $0A, 0B$ stand for the two boundary tetrahedra.}
\end{figure}

To write the spin foam amplitude, it is easier to work in the dual of the triangulation, depicted in the Figure \ref{fig:4-dipole}. It is called the 4-dipole because it has two vertices (of degree 5) connected by four lines, with one external line hanging out of each vertex. The vertices are labeled $A, B$ and the internal lines $1,2,3,4$. The two external lines are denoted $0A$ and $0B$. The lines $1,2,3,4$ create six internal faces, labeled by the pairs of lines, $(ij), 1\leq i<j\leq 4$, since each face goes along two lines only. The four edges shared by the two 4-simplices correspond to four bubbles in $\Gamma$, whose boundary are the faces $(ij), (jk), (ki)$, for $1\leq i<j<k\leq 4$, glued two by two. The surfaces of the bubbles are therefore spherical.

The triangles of the boundary tetrahedra are dual to external faces. As the two 4-simplices share these triangles, the external faces actually go along both $A$ and $B$. There are four external faces, all going along the line $0A$, then choosing an internal line $i=1,2,3,4$, and then all going along the line $0B$. Note that these faces are broken (they are not closed, because the dual triangles are on the boundary of the gluing).

In the spin representation, the amplitude for the 4-dipole has two 10j-symbol, one associated to the vertex $A$ and the other to $B$, and they depend on the spins associated to the faces. Since the corresponding 4-simplices share all their triangles, it means that the spins on the faces are all common to the two 10j-symbols (there are actually only ten faces in the 4-dipole). We thus get the square of a 10j-symbol. Moreover, the spins of the six internal faces must be summed, while the spins on the external (broken) faces are fixed to $j_{i}, i=1,2,3,4$. The partition function for this piece of triangulation is a function of the four external spins,
\be \label{Zdipole-spinrep}
Z_{\rm 4-dipole}(j_1, j_2,j_3, j_4) = \sum_{\substack{j_{ij} \\ 1\leq i<j\leq4}} \left[ \begin{array}{c} \includegraphics[scale=0.4]{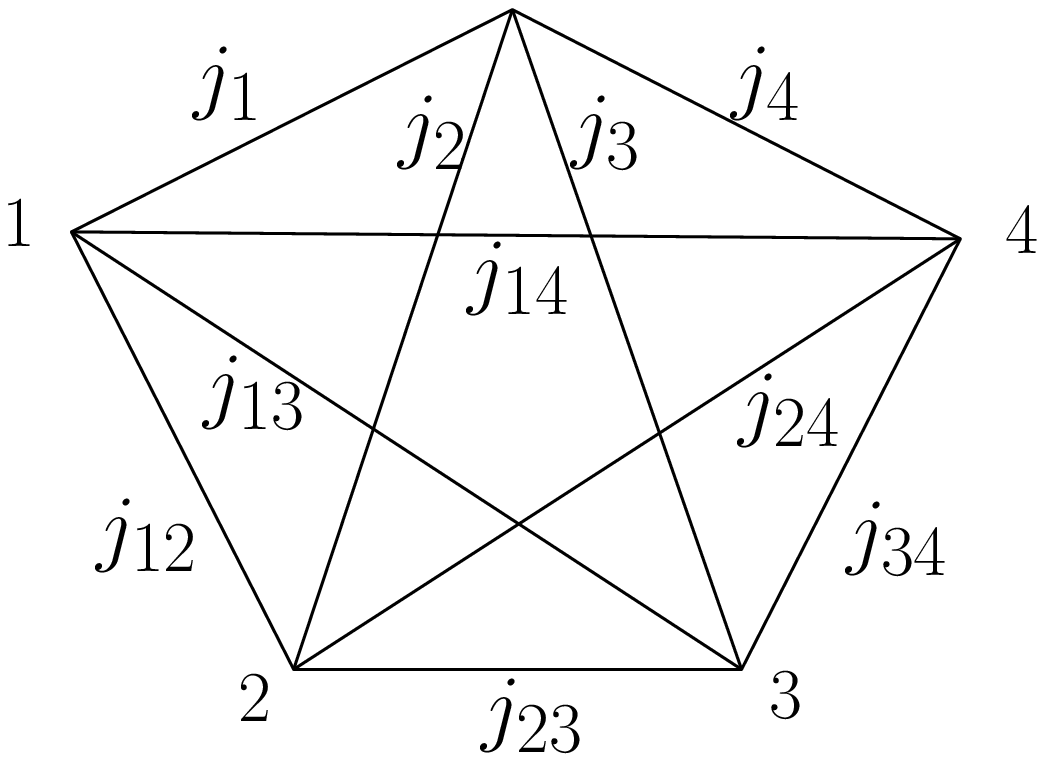} \end{array} \right]^2.
\ee
There is no $d_j$ factors because each face only goes along two lines.

To estimate the potential divergence of these sums, it has been proposed in \cite{simone-spin0} to set the spins of the four external faces to zero. With $j_i=0$, each 10j reduces to the square of a Wigner 6j-symbol (this can be seen in \eqref{10j} which directly reduces to the group integral formulation of the square of the 6j-symbol). Therefore the amplitude is
\be
Z_{\rm 4-dipole}(0,0,0,0) = \sum_{\substack{j_{12}, j_{13}, j_{14}\\ j_{23}, j_{24}, j_{34}}} \begin{Bmatrix} j_{12} &j_{13} &j_{14} \\j_{34} &j_{24} &j_{23}\end{Bmatrix}^4
\ee
We consider the large spin behavior of the summand, when all spins are homogeneously scaled by $\Lambda\gg 1$. The Ponzano-Regge asymptotics of the 6j-symbol states that $\{6j\}\sim 1/\Lambda^{3/2}$. The summand thus behaves like $1/\Lambda^6$ and as there are exactly six sums to perform, it is not possible to conclude about the convergence/divergence of the amplitude. This shows that the choice of measure in our version of the BC model requires a more subtle analysis. However, if gauge symmetries, or redundancies of delta functions in the group integral formulation, were present, a positive exponent of $\Lambda$ would certainly be expected. It means that the simple power-counting argument still suggests the absence of redundancies and gauge symmetries. This is what we will show below.

This is done by moving first to the group integral formulation with effective face weights. The faces $(ij)$, for $1\leq i<j\leq4$, have effective face weights
\be
\omega_{ij} = \int d\gamma_{ij}\ \delta\Bigl(h_{iA}\,h_{jA}^{-1}\ \gamma_{ij}\,h_{jB}\,h_{iB}^{-1}\,\gamma_{ij}^{-1}\Bigr).
\ee
As the external faces are not closed, we cannot use effective face weights. But using the definition \eqref{10j} for the two 10j-symbols, it is easy to see that
\be \label{Zdipole-extfaces}
Z_{\rm 4-dipole}(j_1,j_2, j_3, j_4) = \int \prod_{i<j} d\gamma_{ij} \prod_{0=1}^4 dh_{iA} dh_{iB}\ \prod_{i<j} \delta\Bigl(h_{iA}\,h_{jA}^{-1}\ \gamma_{ij}\,h_{jB}\,h_{iB}^{-1}\,\gamma_{ij}^{-1}\Bigr)\ \prod_{i=1}^4 \chi_{j_i}(h_{0A} h_{iA}^{-1})\,\chi_{j_i}(h_{0B} h_{iB}^{-1}).
\ee
In other words, in the character expansion \eqref{omega characters} of the external faces, only the characters on the wedges of $A$ and $B$ appear. Since there are no delta functions on the external faces (the reason being that we work at fixed external spins), the bubble divergences can only come from the product of delta functions on the internal faces. Therefore, we simplify the analysis by removing the contribution of the external faces, as in the Section \ref{sec:mainassumption}, which here amounts to simply ignore the oscillations of the characters $\chi_{j_i}$ by setting $j_i=0$, like in \cite{simone-spin0}.

The partition function we want to evaluate is
\be \label{Z4-dipole}
Z_{\rm 4-dipole}(0,0,0,0) = \int \prod_{i<j} d\gamma_{ij} \prod_{i=1}^4 dh_{iA} dh_{iB}\ \prod_{i<j} \delta\Bigl(h_{iA}\,h_{jA}^{-1}\ \gamma_{ij}\,h_{jB}\,h_{iB}^{-1}\,\gamma_{ij}^{-1}\Bigr).
\ee
By a redefinition of the elements $h_{iB} \leftarrow h_{iB} h_{1B}^{-1}$ and $h_{iA} \leftarrow h_{iA} h_{1A}^{-1}$, $i=2,3,4$, the elements $h_{1B}, h_{1A}$ are trivially eliminated. Then the faces $(12),(13),(14)$ are used to integrate $h_{iA}$, $i=2,3,4$, which are constrained to be
\be
h_{iA} = \gamma_{1i}\,h_{iB}\,\gamma_{1i}^{-1},\qquad i=2,3,4.
\ee
Only three delta functions remain, which impose constraints between the elements $\gamma_{ij}$ and $h_{2B}, h_{3B}, h_{4B}$. Dropping the subscript $B$,
\begin{multline}
Z_{\rm 4-dipole}(0,0,0,0) = \int dh_{2} dh_3 dh_4 \prod_{i<j} d\gamma_{ij}\ \delta(\gamma_{12} h_2 \gamma_{12}^{-1}\, \gamma_{13}h_3^{-1}\gamma_{13}^{-1}\,\gamma_{23} h_3 h_2^{-1} \gamma_{23}^{-1})\\ \delta(\gamma_{12} h_2 \gamma_{12}^{-1}\, \gamma_{14}h_4^{-1}\gamma_{14}^{-1}\,\gamma_{24} h_4 h_2^{-1} \gamma_{24}^{-1}) \delta(\gamma_{13} h_3 \gamma_{13}^{-1}\, \gamma_{14}h_4^{-1}\gamma_{14}^{-1}\,\gamma_{34} h_4 h_3^{-1} \gamma_{34}^{-1}).
\end{multline}
$\gamma_{12}$ can be completely absorbed into a re-definition of the other $\gamma_{ij} \leftarrow \gamma_{12}^{-1} \gamma_{ij}$. We obtain
\begin{multline} \label{Zdipole}
Z_{\rm 4-dipole}(0,0,0,0) = \int dh_{2} dh_3 dh_4\,d\gamma_{13} d\gamma_{14} d\gamma_{23} d\gamma_{24} d\gamma_{34}\quad \delta(h_2 \, \gamma_{13}h_3^{-1}\gamma_{13}^{-1}\,\gamma_{23} h_3 h_2^{-1} \gamma_{23}^{-1})\\ \delta(h_2\, \gamma_{14}h_4^{-1}\gamma_{14}^{-1}\,\gamma_{24} h_4 h_2^{-1} \gamma_{24}^{-1})\ \delta(\gamma_{13} h_3 \gamma_{13}^{-1}\, \gamma_{14}h_4^{-1}\gamma_{14}^{-1}\,\gamma_{34} h_4 h_3^{-1} \gamma_{34}^{-1}).
\end{multline}

Notice that when a group element appears in one delta, it also appears with its inverse in the same delta. It means that it is now necessary to study the remaining constraints in depth. Let us look at the constraint on the face $(23)$, appearing in the first line of the above Equation. It reads
\be \label{constraint23}
h_2 \, \gamma_{13}h_3^{-1}\gamma_{13}^{-1} = \gamma_{23} h_2 h_3^{-1} \gamma_{23}^{-1}.
\ee
We write $h_i = \cos \theta_i +i\sin\theta_i\,\hat{n}_i\cdot \vec{\sigma}$ and take the trace (in the fundamental representation) on both sides. Using the fact that $\tr (hg) = \cos \theta_h \cos\theta_g - \sin\theta_h\sin\theta_g (\hat{n}_h\cdot \hat{n}_g)$, it becomes when $\theta_2,\theta_3 \neq 0,\pi$,
\be
\hat{n}_2\cdot R(\gamma_{13})\hat{n}_3 = \hat{n}_2\cdot \hat{n}_3,
\ee
where $R(\gamma_{13})$ is the matrix of $\gamma_{13}$ in the 3-dimensional, vector representation. The solution of this equations are easily found for arbitrary $\hat{n}_2,\hat{n}_3$,
\be
\gamma_{13} = e^{-i\phi_{13} \hat{n}_2\cdot\vec{\sigma}}\ e^{i\epsilon\arccos (\hat{n}_2\cdot\hat{n}_3) (\hat{n}_2\times\hat{n}_3)\cdot\vec{\sigma}}\ e^{i\theta_{13} \hat{n}_3\cdot \vec{\sigma}}.
\ee
The angles $\phi_{13},\theta_{13}$ are totally free and parametrize arbitrary rotations around $\hat{n}_2,\hat{n}_3$. Notice that $\epsilon=0,1$ is a discrete ambiguity due to the fact that $\hat{n}_3$ can be rotated in the plane spanned by $\hat{n}_2,\hat{n}_3$ by an angle which is twice the angle between $\hat{n}_2$ and $\hat{n}_3$ without changing the scalar product $\hat{n}_2\cdot\hat{n}_3$. Since it is a discrete ambiguity, we can restrict attention to the case $\epsilon=0$. Inserting this solution in the initial constraint, we find that $h_2 h_3^{-1}$ has to commute with $e^{i\phi_{13} \hat{n}_2\cdot \vec{\sigma}} \gamma_{23}$. Two commuting $\SU(2)$ elements must lie in the same $\U(1)$ sub-group generated by their common rotation axis. As a result,
\be
\gamma_{23} = e^{-i\phi_{13} \hat{n}_2\cdot \vec{\sigma}}\ e^{i\theta_{23} \hat{n}_{23}\cdot\vec{\sigma}},
\ee
where $\hat{n}_{23}$ is the rotation axis of $h_2 h_3^{-1}$. The constraint \eqref{constraint23} thus admits solutions for arbitrary generic $h_2,h_3$, where among the six real degrees of freedom of $\gamma_{13},\gamma_{23}$, only $\phi_{13}, \theta_{13},\theta_{23}$ are left undetermined. It means that three real parameters have been fixed by the three real constraints.

The same reasoning applies to the constraint $h_2 \, \gamma_{14}h_4^{-1}\gamma_{14}^{-1} = \gamma_{24} h_2 h_4^{-1} \gamma_{24}^{-1}$. It leads to solutions for arbitrary $h_2,h_4$ and
\be
\gamma_{14} = e^{-i\phi_{14} \hat{n}_2\cdot\vec{\sigma}}\ e^{i\theta_{14} \hat{n}_4\cdot \vec{\sigma}}, \qquad \gamma_{24} = e^{-i\phi_{14} \hat{n}_2\cdot \vec{\sigma}}\ e^{i\theta_{24} \hat{n}_{24}\cdot\vec{\sigma}},
\ee
up to some discrete ambiguity, where the angles $\phi_{14},\theta_{14},\theta_{24}$ are free. The final step is to insert these solutions into the last set of three real constraints, $\gamma_{13} h_3 \gamma_{13}^{-1}\, \gamma_{14}h_4^{-1}\gamma_{14}^{-1}\,\gamma_{34} h_4 h_3^{-1} \gamma_{34}^{-1}=\mathbbm{I}$, and see whether none of them are trivially satisfied. The following projection of the constraint
\be \label{trace34}
\tr\ \gamma_{13} h_3 \gamma_{13}^{-1}\, \gamma_{14}h_4^{-1}\gamma_{14}^{-1} = \tr\ h_3\,h_4^{-1},
\ee
gives $R(e^{-i\phi_{23} \hat{n}_2\cdot\vec{\sigma}}) \hat{n}_3\cdot R(e^{-i\phi_{14}\hat{n}_2\cdot\vec{\sigma}})\hat{n}_4 = \hat{n}_3\cdot\hat{n}_4$, which is solved by
\be
e^{i(\phi_{13}-\phi_{14}) \hat{n}_2\cdot\vec{\sigma}} = e^{-i\phi_{34}\hat{n}_3\cdot\vec{\sigma}}\ e^{i\psi_{34} \hat{n}_4\cdot\vec{\sigma}},
\ee
for some angles $\phi_{34},\psi_{34}$, up to some discrete ambiguity. For generic $h_2,h_3,h_4$, the rotation axes $\hat{n}_2,\hat{n}_3,\hat{n}_4$ are linearly independent. Therefore, there is no non-trivial solution in the neighbourhood of the trivial solution $\phi_{13}=\phi_{14}, \phi_{34}=\psi_{34}=0$. It means that \eqref{trace34} indeed removes one degree of freedom by setting $\phi_{13}=\phi_{14}$. Finally, the constraint imposes that $h_3 h_4^{-1}$ commutes with $e^{i\phi_{13}\hat{n}_2\cdot\vec{\sigma}} \gamma_{34}$. This fixes the two real degrees of freedom of the rotation axis of $e^{i\phi_{13}\hat{n}_2\cdot\vec{\sigma}} \gamma_{34}$ and leaves one angle, denoted $\theta_{34}$, free,
\be
\gamma_{34} = e^{-i\phi_{13}\hat{n}_2\cdot\vec{\sigma}}\ e^{i\theta_{34}\hat{n}_{34}\cdot\vec{\sigma}},
\ee
where $\hat{n}_{34}$ is the rotation axis of $h_3h_4^{-1}$. The last set of constraints has thus eliminated three real parameters, meaning that we did not meet any redundancies while solving the constraints.

The parameters left undetermined are $h_2,h_3,h_4, \phi_{13}, \theta_{13}, \theta_{23}, \theta_{14},\theta_{24},\theta_{34}$, and the set of solutions is
\begin{multline}
\cF = \left\{h_2,h_3,h_4,\gamma_{13} = k_2\,e^{i\theta_{13}\hat{n}_{3}\cdot\vec{\sigma}}, \gamma_{14} = k_2\,e^{i\theta_{14}\hat{n}_{4}\cdot\vec{\sigma}}, \gamma_{23} = k_2\,e^{i\theta_{23}\hat{n}_{23}\cdot\vec{\sigma}}, \gamma_{24} = k_2\,e^{i\theta_{24}\hat{n}_{24}\cdot\vec{\sigma}}, \gamma_{34} = k_2\,e^{i\theta_{34}\hat{n}_{34}\cdot\vec{\sigma}}; \right.
\\ \left. \text{with $h_2, h_3, h_4$ arbitrary and $k_2 = e^{i\phi_{13}\hat{n}_2\cdot\vec{\sigma}}$} \right\}.
\end{multline}
This is a fifteen-dimensional space, while there were $8$ $\SU(2)$ elements to integrate in \eqref{Zdipole}, i.e. $8\times 3=24$ real variables. Therefore, the constraints have put restrictions on $24-15=9$ variables, corresponding to the number of constraints.

This means that there are no divergences coming from the product of delta functions. Defining the map $H: \SU(2)^8 \to \SU(2)^3$
\begin{multline}
H(h_2, h_3, h_4, \gamma_{13}, \gamma_{14}, \gamma_{23},\gamma_{24}, \gamma_{34}) \\= \left(h_2 \, \gamma_{13}h_3^{-1}\gamma_{13}^{-1}\,\gamma_{23} h_3 h_2^{-1} \gamma_{23}^{-1},\ h_2 \, \gamma_{14}h_4^{-1}\gamma_{14}^{-1}\,\gamma_{24} h_4 h_2^{-1} \gamma_{24}^{-1},\ \gamma_{13} h_3 \gamma_{13}^{-1}\, \gamma_{14}h_4^{-1}\gamma_{14}^{-1}\,\gamma_{34} h_4 h_3^{-1} \gamma_{34}^{-1}\right),
\end{multline}
the above analysis reveals that for a generic solution $\phi\in\cF$, $\dim \ker dH_\phi = 15$ and $\rk dH_\phi = 9$. Therefore the tangent space at $\phi$ decomposes as $T_\phi\SU(2)^8 = T_\phi\cF \oplus N_\phi\cF$, where the normal space $N_\phi\cF$ is the ortho-complement of the tangent space to the space of solutions. The restriction $dH_{\phi|N_\phi\cF}$ to the normal space is an invertible map from $N_\phi\cF$ to $T_{(\mathbbm{I}, \mathbbm{I}, \mathbbm{I})}\SU(2)^3$. The integral over the normal directions correspond to the parameters which are fixed by constraints. The partition function becomes
\be
Z_{\rm 4-dipole}(0,0,0,0) = \int_{\cF} d\phi\ \frac1{|\det dH_{\phi|N_\phi\cF} |}.
\ee
For generic solution $\phi$, this determinant is non-vanishing, but it may happen that it actually vanishes on a subset of measure zero in $\cF$. Such singularities are well-known to arise in BF theory \cite{homology-bf}. In two-dimensional BF, there \emph{are} such singularities on the moduli space of flat connections \cite{sengupta-singularities-2d}, but it can be shown that these are integrable (as expected since the partition function can be exactly calculated using the character expansion as $\zeta(2g-2)$ on a surface of genus $g$). However, there is absolutely no generic result on such singularities beyond the two-dimensional case, meaning that one has to deal with them case by case. In the present case, this is quite complicated because of the number of variables involved. We will not discuss further the possibility that the integral is divergent due to such singularities. Instead, we conclude this Section by emphasizing the fact that there are no redundancies in the initial product of delta functions, removing the expected source of divergences in spin foams.

\section{Generic evaluation of the BC partition function}

The technique used in the last section was developed in \cite{homology-bf} to integrate over the representation variety of finitely presented fundamental groups. It can also be applied to the BC model on generic 2--complexes.

Combining \eqref{hol rep} with \eqref{BCfaceweight}, we see that the partition function writes as an integral over
\be
\mathcal{A} = \Bigl\{ A = \bigl(\{h_{ev}\}_{e\in\Gamma_1, v\in\Gamma_0}, \{\gamma_{vf}\}_{v\in\gamma_0, f\in\Gamma_2}\bigr) \Bigr\},
\ee
subjected to the constraint
\be
H(A) \equiv \left\{ H_f = \gamma_{v_1 f} h_{e_1 v_1} h_{e_2 v_1}^{-1} \gamma_{v_1 f}^{-1} \dotsm \gamma_{v_n f} h_{e_n v_n} h_{e_1 v_n}^{-1} \gamma_{v_n f}^{-1} \right\}_{f\in\Gamma_2} = \mathbbm{I},
\ee
where $v_1,\dotsc,v_n$ and $e_1,\dotsc,e_n$ are the vertices and lines around the boundary of each face $f\in\Gamma_2$. $H$ is a map from $\mathcal{A}$ to $\SU(2)^{|\Gamma_2|}$ and
\be
Z(\Gamma) = \int_{\mathcal{A}} dA\quad \delta\bigl(H(A)\bigr).
\ee
$\delta(H)$ is the $3|\Gamma_2|$-dimensional delta function over the target space $\SU(2)^{|\Gamma_2|}$, $\delta(H) = \prod_f \delta(H_f)$. Thinking of the group elements as holonomies of a gauge field, and of $H_f$ as the corresponding Wilson loops, the integral corresponds to a lattice gauge theory on a two-complex at zero coupling. The integral localizes on the set $\cF = H^{-1}(\mathbbm{I})$ (which is the set of flat connections in the lattice gauge theory interpretation).

Therefore, we have to solve the constraints, meaning that we need to find among the real degrees of freedom of $A\in\mathcal{A}$ those which are free and parametrize the set of solutions $\cF$, and those which are functions of the free parameters as determined by the constraints. Because $(H_f=\mathbbm{I})_{f\in \Gamma_2}$ is a set of polynomial equations on $\SU(2)$, the set of solutions $\cF$ is a real algebraic variety whose dimension is the number of free parameters. For generic solutions $\phi\in \cF$, this turns out to coincide with the dimension of the kernel of $dH_\phi$. This is because
\be
T_\phi\cF =\ker dH_\phi,
\ee
as expected. 

Now we can understand the potential divergences coming from the product of delta functions. Notice that the rank of $dH_\phi$ is $\rk dH_\phi = 3|\Gamma_2|-\ker dH_\phi$ and corresponds to the number of directions spanned by $dH_\phi$ in the target space $\SU(2)^{|\Gamma_2|}$. If $\rk dH_\phi$ is strictly less than $\dim \SU(2)^{|\Gamma_2|}=3|\Gamma_2|$ for generic $\phi$, then it means that some directions are not explored whatever the variations around $\phi$ are. Therefore the components of the delta functions along these directions are trivially satisfied and the amplitude is divergent.

It is actually possible to describe the divergence rate, following \cite{homology-bf, bubblediv3}. If the delta functions are regularized with a thin width $1/\Lambda$ (using a heat kernel for instance), then the divergence degree is $\Lambda^{3|\Gamma_2| - \rk dH_\phi}$. In the case of BF theory on a 2-complex $\Gamma$, it is possible to relate this divergence rate to the topology of $\Gamma$ (and even to the spacetime topology if $\Gamma$ is the 2-skeleton of a cell decomposition of a four-dimensional manifold). However, in the Barrett-Crane, we have not found a simple topological interpretation of the divergence rate $3|\Gamma_2| - \rk dH_\phi$.

When $\rk dH_\phi$ is exactly the dimension of the target space for generic $\phi$, one can conclude that there is no divergence coming from the product of the delta functions, i.e. all the constraints are independent. This was the case for the 4-dipole. Moreover, the integral $Z(\Gamma)$ rewrites as an integral over $\cF(\Gamma)$, and for each $\phi\in\cF(\Gamma)$, an integral over the directions orthogonal to $T_\phi \cF(\Gamma)$, denoted $N_\phi\cF$ for `normal space'. The directions of the normal space are those along which the constraints fix the variations around $\phi$ to vanish. Therefore, the integral becomes
\be \label{BC localized}
Z(\Gamma) = \int_{\cF(\Gamma)} d\phi \int_{N_\phi\cF} da\ \delta(dH_\phi(a)) = \int_{\cF(\Gamma)} d\phi\ \frac{1}{\lvert \det dH_{\phi|N_\phi\cF}\rvert}.
\ee
$dH_{\phi|N_\phi\cF}$ is the restriction of $dH_\phi$ to the normal space (intuitively, its kernel has been removed).

\section{Some gauge symmetries and recursion relations on the 10j-symbol}\label{sec:recurs}

\subsection{Existence of gauge symmetries at certain solutions}

Gauge symmetries around a solution $\phi\in\cF$ correspond to directions which are not spanned by $dH_\phi$. To see their action, let us re-write the $\delta(dH_\phi(a))$ of \eqref{BC localized} as
\be\label{71}
\delta( dH_\phi(a)) = \int_{\su(2)^{|\Gamma_2|}}db\ \exp i\langle b, dH_\phi(a) \rangle,
\ee
where $\langle \cdot,\cdot \rangle = \sum_f \langle \cdot, \cdot \rangle_f$ is the sum of the invariant inner product over the different copies of $\su(2)$. The variable $b=\{b_f\}_{f\in\Gamma_2}$ is a Lagrange multiplier imposing the constraint. A gauge symmetry is a $\phi$-dependent, non-zero variation $\delta_\phi b$ which leaves the action $\langle b, dH_\phi(a) \rangle$ invariant,
\be \label{def gauge}
\langle \delta_\phi b,dH_\phi(a)\rangle = 0,
\ee
for any variation $a\in T_\phi\mathcal{A}$. Typically, we want a gauge symmetry to involve non-trivial $\delta_\phi b_f\neq 0$ on several faces\footnote{ \label{foot:torus}In BF theory on the torus,
\be
Z_{\text{BF 2-torus}} = \int_{\SU(2)^2} dx\,dy\ \delta(xyx^{-1}y^{-1}),
\ee
the constraint forces $x$ and $y$ to lie in the same $\U(1)$ subgroup, say generated by $\vec{\sigma}\cdot \hat{n}$. Then it is quite easy to see that the linearized constraint $d(xyx^{-1}y^{-1})$ never spans the direction $\vec{\sigma}\cdot \hat{n} \in\su(2)$. However, this is clearly not a phenomenon we want to call gauge symmetry.}.

When the constraints are independent, $dH_\phi$ has maximal rank and there is generically no gauge symmetry. We have seen this is the case for the 4-dipole, and single bubble contributions are finite so they have no gauge symmetry either. This is expected in theory of discretized gravity such as spin foams, because going on the lattice breaks diffeomorphism invariance \cite{broken-sym}. There is however a  special class of triangulations which only admit flat solutions and for which one would expect diffeomorphism symmetry also in the discrete case \cite{DittrichRyan}.
%
Physically, the gauge symmetry in the flat space case means that vertices of the triangulation can be moved around without changing the physics. 
Such a vertex translation symmetry arises for instance for the 5-1 move configuration (arising from a subdivision of a 4--simplex into 5 simplices), where the inner vertex can be moved in four directions without changing the flatness of the configuration.
While these symmetries have been observed and (canonically) analyzed in details in Regge calculus \cite{hoehn} and shown to lead to Dirac's hypersurface deformation algebras \cite{vbbd}, the (quantum) spin foam case remains mostly unexplored (with the exception of 3D gravity and BF theory in higher dimensions \cite{cellular-bf, wdw3d, recursion15j}. Assuming such symmetries would exist, we could derive Hamiltonian constraint operators for the boundary wave functions defined by the spin foam transition amplitudes. This would lead to a canonical theory describing the amplitudes defined by spin foams and thus to a connection between canonical LQG and spin foams. The case of BF theory, where the symmetries are not broken, has been discussed in \cite{wdw3d,recursion,recursion15j}. For the BC model we found only one configuration where such a symmetry occurred - the case of two two--valent faces glued to each other. Indeed this constitutes the only bubble divergence we found. Hence we do not expect a divergence (and hence no full symmetry) for the 5-1 move configuration.

Additionally to the special class of triangulations which only support flat solutions, such vertex translation symmetries might arise around special (i.e. flat) solutions in more general triangulations. The Hessian evaluated on such solutions will have zero modes corresponding to infinitesimal vertex translation symmetry \cite{williams}. Again such symmetries have not been discussed in the spin foam case yet. Here we ask whether a possible similar phenomenon exist for spin foams, i.e. symmetries which occur only at special solutions.

Indeed we will show that some solutions $\phi\in\cF$ to the constraint have gauge symmetries. We have already mentioned that even when the constraints are independent for generic solutions, there may be some solutions $\phi$ (a set of measure zero in $\cF$) which are singular because $\det dH_{\phi|N_\phi\cF} = 0$, meaning that $\ker dH_\phi$ becomes larger than $T_\phi\cF$ and not all directions of the target space $\SU(2)^{|\Gamma_2|}$ are spanned. Some of these singularities may be interpreted as the appearance of gauge symmetries \eqref{def gauge} (but not all singularities correspond to gauge symmetries, since the phenomenon of footnote \ref{foot:torus} can also occur in singularities). 

The condition \eqref{def gauge} has to hold for any variation, in particular when applied to a variation $\xi_{v_i f}$ of a group element $\gamma_{v_i f}$,
\be
\langle \delta_\phi b_f, \Ad(\gamma_{v_1 f} h_{e_1 v_1} h_{e_2 v_1}^{-1} \gamma_{v_1 f}^{-1} \dotsm \gamma_{v_{i-1} f} h_{e_{i-1} v_1} h_{e_i v_1}^{-1} \gamma_{v_{i-1} f}^{-1}) \Bigl[ 1 - \Ad(\gamma_{v_i f} h_{e_i f} h_{e_{i+1} f}^{-1} \gamma_{v_i f}^{-1}) \Bigr] \xi_{v_i f} \rangle =0.
\ee
There are two obvious situations where this is true:
\begin{itemize}
\item when all the group elements $h_{ev}$ are the same around each vertex $h_{ev}=h_v$, because the operator $1 - \Ad(\gamma_{v_i f} h_{e_i f} h_{e_{i+1} f}^{-1} \gamma_{v_i f}^{-1})$ is then identically zero, for any elements $\gamma_{vf}$, (below we will need in addition that the elements $\gamma_{vf}$ only depend on the face),
\item when all group elements $\gamma_{vf}, h_{ev}$ lie in the same $\U(1)$ sub-group of $\SU(2)$, say generated by $\vec{\sigma}\cdot \hat{n}$, and $\delta_\phi b$ are variations in this direction. This is because $[1-\Ad(e^{i\alpha \hat{n}\cdot\vec{\sigma}})](\hat{n}\cdot\vec{\sigma}) = 0$.
\end{itemize}

In both cases, we can exhibit gauge symmetries, when $\Gamma$ is a $d$-dimensional cell complex, with $d\geq 3$, using cellular homology. The chain spaces $C_i(\Gamma)$ are real vector spaces, $C_i \simeq \R^{|\Gamma_i|}$ (hence we identify chains and co-chains), with boundary operators $\partial_i$ and co-boundary operators $\delta^i$,
\be
0 \rightleftarrows\ C_0(\Gamma)\ \overset{\delta^0}{\underset{\partial_1}\rightleftarrows}\ C_1(\Gamma)\ \overset{\delta^1}{\underset{\partial_2}\rightleftarrows}\ C_2(\Gamma)\ \overset{\delta^2}{\underset{\partial_3}\rightleftarrows}\ \dotsb \overset{\delta^{d-1}}{\underset{\partial_d}\rightleftarrows}\ C_d(\Gamma)\ \rightleftarrows 0,\\
\ee
The boundary and co-boundary operators satisfy $\partial_{i-1}\circ \partial_i \equiv0$ and $\delta^{i}\circ \delta^{i-1} \equiv0$ and they are dual to each other.

The operator $\delta^{1}$ sends lines to faces: if $v=\{v_e\}_{e\in\Gamma_1}$ then $\delta^1(v) = \{\delta^1(v)_{|f}\}_{f\in\Gamma_2}$ with
\be
\delta^1 (v)_{|f} = \sum_{e\subset f} \epsilon_{ef}\ v_e,
\ee
where $\epsilon_{ef}=\pm$ denotes the relative orientation between $e$ and $f$.

In the case the elements $h_{ev}$ only depend on the vertices,
\be \label{special conf1}
h_{ev}=h_v, \qquad \text{and}\qquad \gamma_{vf}=\gamma_f,
\ee
on the faces, the differential of $H$ reduces to
\be
\begin{aligned}
dH_{\{h_{ev}=h_v, \gamma_{vf}=\gamma_f\}| f} &= \Ad(\gamma_f) \Bigl(dh_{e_1 v_1} h_{e_1 v_1}^{-1} - dh_{e_2 v_1} h_{e_2 v_1}^{-1} + \dotsb +dh_{e_n v_n} h_{e_n v_n}^{-1} - dh_{e_1 v_n} h_{e_1 v_n}^{-1}\Bigr),\\
&= -\Ad(\gamma_f) \sum_{e\subset f} \epsilon_{ef}\ \bigl( dh_{e s(e)} h_{e s(e)}^{-1} - dh_{e t(e)} h_{e t(e)}^{-1}\bigr).
\end{aligned}
\ee
Here $s(e), t(e)$ denote the source and target vertices of the line $e$. Therefore, $\delta^1$ and $dH$ are simply related. Denote $\Theta_h = dh h^{-1}: T_h\SU(2) \to \su(2)$ the Maurer-Cartan form which maps the tangent space at $h$ to the Lie algebra. It becomes
\be
dH_{\{h_{ev}=h_v, \gamma_{vf}=\gamma_f\}|f}(\{a_{ev}\}) = -\delta^1_f \otimes \Ad(\gamma_f) \Theta_{h_v}\bigl( a_{es(e)} - a_{e t(e)}\bigr),
\ee
for any tangent vectors $a_{ev}\in T_{h_v}\SU(2)$. The adjoint action by $\gamma_f$ on each face can be absorbed into a re-definition of the Lagrange multipliers $b_f\leftarrow \Ad(\gamma_f^{-1}) b_f$, so that $dH$ is basically the cellular co-boundary operator $\delta^1$. We notice that
\be
\langle b+\partial_3\otimes \id_{\su(2)}(c), dH_\phi(a) \rangle = \langle b, dH_\phi(a)\rangle + \langle c, \bigl(\delta^2\otimes \id_{\su(2)}\bigr) \circ dH_\phi(a)\rangle,
\ee
using $\langle \partial_3\otimes \id_{\su(2)}(c), x\rangle = \langle c, \delta^2\otimes \id_{\su(2)}(x)\rangle$. Thanks to the identity $\delta^2 \circ \delta^1 \equiv 0$, we see that
\be
b\mapsto b+\partial_3\otimes \id_{\su(2)} (c),
\ee
for any $c \in C_3(\Gamma)\otimes \su(2)$ is a gauge transformation. This is the same gauge symmetry as in the topological BF theory with structure group $\R^3$. The reason is that the solution we are looking at is up to local rotations (at the vertices) equivalent to the trivial solution where all group elements are the identity, and in the neighborhood of the identity, $\SU(2)$ looks like $\R^3$.

In the case where all group elements are generated by a single direction, $\vec{\sigma}\cdot \hat{n}$, the gauge symmetry is the same as in a $\U(1)$ BF theory. Let us parametrize the group elements as
\be
h_{ev} = e^{i\theta_{ev}\, \hat{n}\cdot\vec{\sigma}},\qquad \gamma_{vf} = e^{i\alpha_{vf}\,\hat{n}\cdot\vec{\sigma}}.
\ee
The constraint $H_f(A)=\mathbbm{I}$ then reduces to a $\U(1)$ constraint,
\be
\sum_{e\subset f} \theta_{e s(e)} - \theta_{e t(e)} = 0 \mod(2\pi),
\ee
making the contact with $\U(1)$ BF theory obvious. The privileged direction $\vec{\sigma}\cdot \hat{n}$ induces a natural splitting of $\su(2) = \u(1)_{\hat{n}} \oplus \alg_\perp$ where $\u(1)_{\hat{n}} = \operatorname{span}\{\vec{\sigma}\cdot \hat{n} \}$ and $\alg_\perp$ is its ortho-complement. This also gives a natural basis in $T_{h_{ev}}\SU(2)$,
\be
T_{h_{ev}}\SU(2) = \R \partial_{\theta_{ev}} \oplus \alg_{ev\perp},
\ee
where $\alg_{ev\perp}$ is spanned by the derivatives with respect to the two components of the rotation axis $\hat{n}_{ev}$ of $h_{ev}$ evaluated at $\hat{n}_{ev}=\hat{n}$. A similar decomposition $T_{\gamma_{vf}}\SU(2) = \R \partial_{\alpha_{vf}} \oplus \alg_{vf\perp}$ holds.

Some straightforward algebra shows that $dH$ sends $\partial_{\theta_{ev}}$ to $\u(1)_{\hat{n}}$, and $\alg_{ev\perp}$ as well as $\alg_{vf\perp}$ to $\alg_\perp$, and that it vanishes on $\partial_{\alpha_{vf}}$. Moreover, the restriction of $dH$ to the sub-spaces $\partial_{\theta_{ev}}$ basically reduces to the cellular co-boundary operator $\delta^1$,
\be
dH_{f}(\{ x_{ev}\partial_{\theta_{ev}}\}) = -\delta^1_f(\{x_{es(e)}-x_{e t(e)}\}) \otimes \vec{\sigma}\cdot\hat{n}.
\ee
Therefore, the action is left invariant by the transformation
\be
b\mapsto b+\partial_3\otimes \id_{\su(2)}(c), \qquad \forall\, c\in C_3(\Gamma)\otimes\u(1)_{\hat{n}}.
\ee

Since these gauge symmetries rely on cellular homology, they are reducible as soon as $d\geq 4$. Indeed, the gauge parameters $c$ are not independent. If two of them differ by $\partial_4(y)$ for $y\in C_4(\Gamma)\otimes \su(2)$ in the first case and $y\in C_4(\Gamma)\otimes \u(1)_{\hat{n}}$ in the second case, then they induce exactly the same gauge transformation (because $\partial_3\circ\partial_4=0$). This reducibility is well-known in BF theory \cite{cellular-bf}.

\subsection{Recursion relations on the 10j-symbol}

\subsubsection{Using the 4-dipole}

We consider the 4-dipole configuration as in the Section \ref{sec:4-dipole} where we wrote the partition function \eqref{Zdipole-extfaces} with fixed spins on the external faces. However, in this partition function, not all solutions are of the form \eqref{special conf1}. Further, we have seen that there is no gauge symmetry for generic solutions since there is no redundancies in the constraints. Therefore, instead of the partition function $Z_{\rm 4-dipole}$ in \eqref{Zdipole-extfaces}, we will consider the following quantity,
\begin{multline} \label{Idipole}
I_{\rm 4-dipole}(j_{1B},j_{2B}, j_{3B}, j_{4B}, j_{23}, j_{24},j_{34}) = \int \prod_{i=0}^4 dh_{iA} dh_{iB} \prod_{1\leq i<j\leq 4} d\gamma_{ij}\ \prod_{i=1}^4 \delta(h_{0A} h_{iA}^{-1})\ \chi_{j_{iB}}(h_{0B} h_{iB}^{-1})\\ \prod_{j=2,3,4}\delta\bigl(h_{1A} h_{jA}^{-1}\gamma_{1j} h_{jB} h_{1B}^{-1}\gamma_{1j}^{-1}\bigr) \prod_{2\leq i<j\leq 4} \chi_{j_{ij}}(h_{iA} h_{jA}^{-1}\gamma_{ij} h_{jB} h_{iB}^{-1}\gamma_{ij}^{-1}\bigr),
\end{multline}
and proceed to evaluate $I_{\rm 4-dipole}$ in two different ways to get recursion relations on the 10j-symbol, which can be interpreted as a (constraint) equation on the vertex amplitude.

But first let us point out the difference between $Z_{\rm 4-dipole}$ and $I_{\rm 4-dipole}$. First, note that the integration variables are the same, only the integrands and the boundary variables differ. We have changed by hand the characters of the external faces at vertex $A$ in \eqref{Zdipole-extfaces} like
\be \label{constraintA}
\prod_{i=1}^4 \chi_{j_i}(h_{0A} h_{iA}^{-1}) \quad \rightarrow \quad \prod_{i=1}^4 \delta(h_{0A} h_{iA}^{-1}) = \sum_{\substack{j_{1A}, j_{2A}\\ j_{3A}, j_{4A}}} \prod_{i=1}^4 d_{j_{iA}}\ \chi_{j_{iA}}(h_{0A} h_{iA}^{-1}).
\ee
This allows to satisfy the special condition \eqref{special conf1} on the elements $h_{ev}$ at the vertex $A$. Notice that the delta functions have an expansion onto characters similar to the initial characters of \eqref{Zdipole-extfaces}. However, in $Z_{\rm 4-dipole}$, the spins of the characters $\chi_{j_i}(h_{0A} h_{iA}^{-1})$ at the vertex $A$ and of the characters $\chi_{j_i}(h_{0B} h_{iB}^{-1})$ at the vertex $B$ are the same, because they correspond to the same (external) faces going along both $A$ and $B$. When putting in $I_{\rm 4-dipole}$ some additional constraints, the equality $j_{iA} = j_{iB} = j_i$ is broken because more modes are necessary to enforce the condition \eqref{special conf1} at vertex $A$.

The dipole possesses six internal faces. Combining the constraint \eqref{constraintA} with the effective face weights of the faces $(1i)$, for $i=2,3,4$, imposes in turn that $h_{iB}=h_{jB}$, for any $1\leq i<j\leq 4$. Indeed,
\be \label{constraintAB}
\delta(h_{0A} h_{1A}^{-1})\ \delta(h_{0A} h_{iA}^{-1})\ \int d\gamma_{1i}\ \delta\bigl(h_{1A} h_{iA}^{-1}\gamma_{1j} h_{iB} h_{1B}^{-1}\gamma_{1i}^{-1}\bigr) = \delta(h_{0A} h_{1A}^{-1})\ \delta(h_{0A} h_{iA}^{-1})\ \delta(h_{iB}\,h_{1B}^{-1}).
\ee
As for the three remaining faces $(23), (24), (34)$, the delta functions of their effective face weights is automatically satisfied thanks to \eqref{constraintAB}, i.e. $h_{iA} h_{jA}^{-1} \gamma_{ij} h_{jB} h_{iB}^{-1} \gamma_{ij}^{-1} = \mathbbm{I}$ (for any $\gamma_{ij}$). Therefore, these delta functions become redundant, confirming in this case the existence of gauge symmetries. To avoid the divergences associated to these redundancies, observe that we have not included the effective face weights of the faces $(23), (24), (34)$ in $I_{\rm 4-dipole}$. Instead, we have only picked up one mode of their character expansion (the last line of products in \eqref{Idipole}). Due to \eqref{constraintAB}, these characters simply evaluates to the dimension of their representation,
\be
\prod_{i=1}^4 \delta(h_{0A} h_{iA}^{-1})\ \delta(h_{iB}\,h_{1B}^{-1})\ \prod_{2\leq i<j\leq 4} \chi_{j_{ij}}(h_{iA} h_{jA}^{-1}\gamma_{ij} h_{jB} h_{iB}^{-1}\gamma_{ij}^{-1}\bigr) = \prod_{i=1}^4 \delta(h_{0A} h_{iA}^{-1})\ \delta(h_{iB}\,h_{1B}^{-1})\ \prod_{2\leq i<j\leq 4} d_{j_{ij}}.
\ee
This product of dimensions is the sole dependence of $I_{\rm 4-dipole}$ in the spins $j_{23}, j_{24}, j_{34}$. If these spins were summed (with measure $d_{j_{ij}}$) to form the effective face weight as in $Z_{\rm 4-dipole}$, we would get $\sum_{j_{ij}} d_{j_{ij}}^2$ which is obviously divergent (it is the formal expansion of $\delta(\mathbbm{I})$). The fact that the dependence of $I_{\rm 4-dipole}$ on $j_{ij}$ ($2\leq i<j\leq 4$) is just $d_{j_{ij}}$ is the signature of the gauge symmetry, similarly to the case of spherical bubbles in BF theory \cite{recursion}

Therefore, the only non-trivial contribution to $I_{\rm 4-dipole}$ is the product of characters on the external faces at the vertex $B$. With the change of variable $h = h_{0B} h_{iB}^{-1}$ (this quantity is independent of $i=1,2,3,4$), we finally get
\be \label{Idipole-lhs}
I_{\rm 4-dipole} = d_{j_{23}}\,d_{j_{24}}\,d_{j_{34}}\ \int dh\ \prod_{i=1}^4 \chi_{j_{iB}}(h),
\ee

The second way to evaluate $I_{\rm 4-dipole}$ is through a character expansion of all the delta functions, and integrating the variables $\gamma_{ij}$. That leads to
\begin{multline} \label{Idipole-rhs}
I_{\rm 4-dipole} = \sum_{\substack{j_{1A}, j_{2A}, j_{3A}, j_{4A}\\ j_{12}, j_{13}, j_{14}}} \frac{\prod_{i=1}^4 d_{j_{iA}}}{d_{j_{23}}\,d_{j_{24}}\,d_{j_{34}}} \left[\int \prod_{i=0}^4 dh_{iA}\ \prod_{1\leq i<j\leq4} \chi_{j_{ij}}(h_{iA} h_{jA}^{-1}) \prod_{i=1}^{4} \chi_{j_{iA}}(h_{0A} h_{iA}^{-1})\right]\\
\times \left[\int \prod_{i=0}^4 dh_{iB}\ \prod_{1\leq i<j\leq4} \chi_{j_{ij}}(h_{iB} h_{jB}^{-1}) \prod_{i=1}^{4} \chi_{j_{iB}}(h_{0B} h_{iB}^{-1})\right].
\end{multline}
The two quantities into square brackets are 10j-symbols, according to the definition \eqref{10j}. Equating this formula with \eqref{Idipole-lhs} leads to
\begin{multline} \label{recursion1}
\sum_{j_{12}, j_{13}, j_{14}} \left[ \sum_{\substack{j_{1A}, j_{2A}\\ j_{3A}, j_{4A}}} \left[\prod_{i=1}^4 d_{j_{iA}}\right]
\begin{array}{c} \includegraphics[scale=0.4]{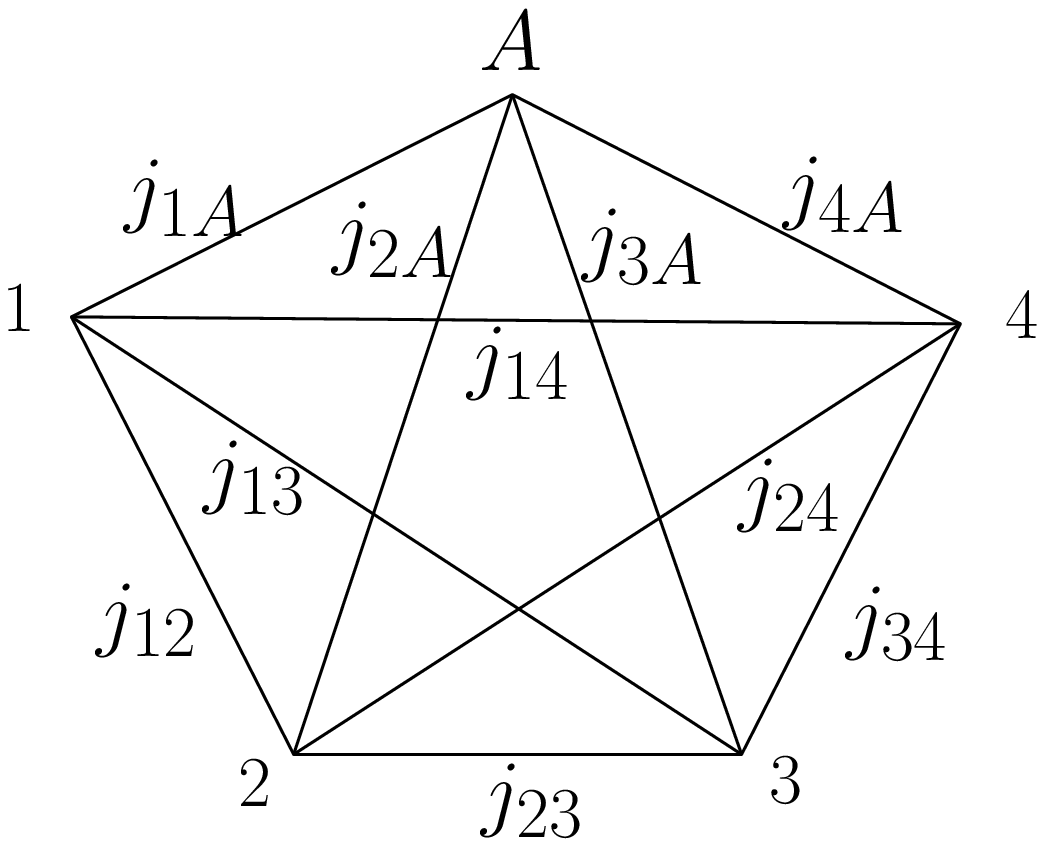} \end{array} \right] \left[ \begin{array}{c} \includegraphics[scale=0.4]{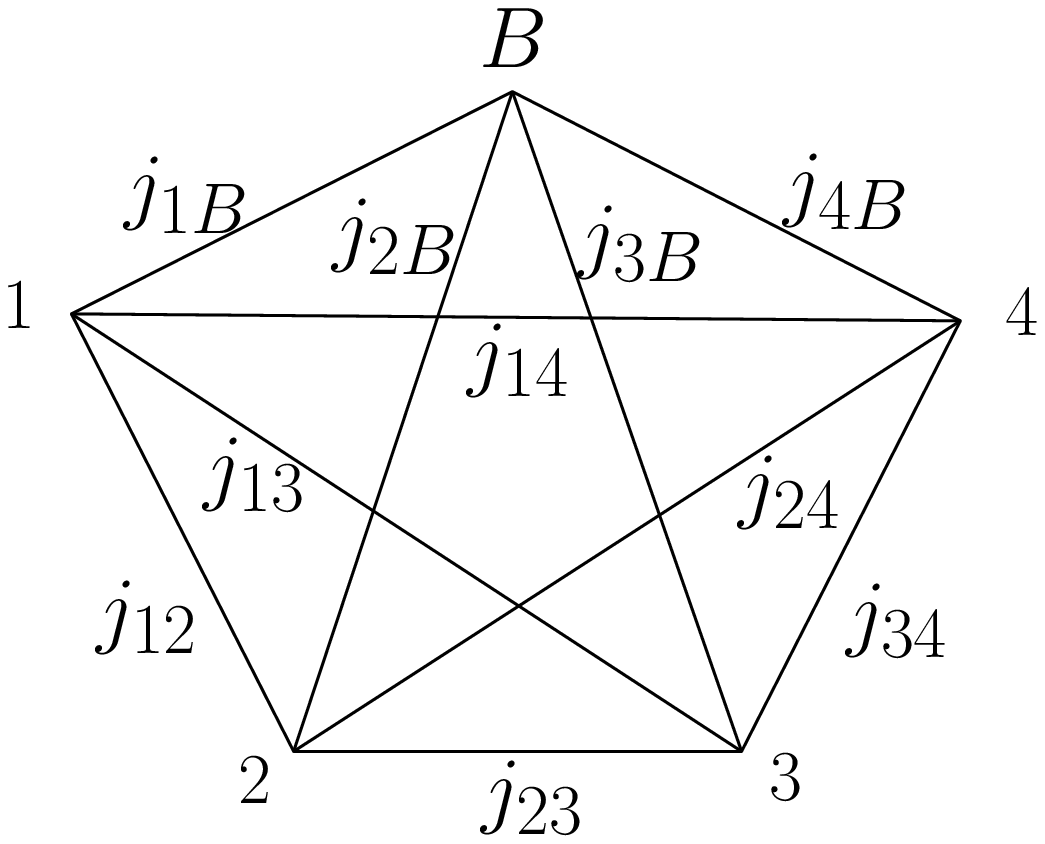} \end{array} \right] \\
= \left[d_{j_{23}}\,d_{j_{24}}\,d_{j_{34}}\right]^2\ \int dh\ \prod_{i=1}^4 \chi_{j_{iB}}(h).
\end{multline}
This is a new sum rule for 10j-symbols. The summand is not the square of the 10j-symbol as in the spin foam representation \eqref{Zdipole} of $Z_{\rm 4-dipole}$ because of the constraints \eqref{constraintA} which implies $j_{iA} \neq j_{iB}$. Moreover, the spins $j_{iA}$ are summed here while the boundary spins $j_i$ in $Z_{\rm 4-dipole}$ are fixed. The final difference is that the spins $j_{23}, j_{24},j_{34}$ are arbitrary but fixed, instead of being summed, to avoid the divergences due to the gauge symmetries.

Our formula can be specialized to specific values of the spins free spins $j_{iB}, i=1,2,3,4, j_{23}, j_{24}, j_{34}$. For instance, setting $j_{iB}=0$, the 10j-symbol in the second bracket collapses to a squared 6j-symbol, and the integral on the right hand side is trivialized as $\chi_{0}(h)=1$,
\be
\sum_{j_{12}, j_{13}, j_{14}} \begin{Bmatrix} j_{12} &j_{13} &j_{14}\\ j_{34} &j_{24} &j_{23} \end{Bmatrix}^2
\left[ \sum_{\substack{j_{1A}, j_{2A}\\ j_{3A}, j_{4A}}} \left[\prod_{i=1}^4 d_{j_{iA}}\right] \begin{array}{c} \includegraphics[scale=0.4]{10jA.eps} \end{array} \right]
= \left[d_{j_{23}}\,d_{j_{24}}\,d_{j_{34}}\right]^2.
\ee

Another interesting way to use our main formula \eqref{recursion1} is to sum over one of the spins $j_{iB}$, say $j_{1B}$, with measure $d_{j_{1B}}$. Then the integral on the right hand side simplifies,
\be
\sum_{j_{1B}} d_{j_{1B}} \int dh\ \prod_{i=1}^4 \chi_{j_{iB}}(h) = \int dh\ \delta(h)\ \chi_{j_{2B}}(h)\ \chi_{j_{3B}}(h)\ \chi_{j_{4B}}(h) = d_{j_{2B}}\,d_{j_{3B}}\,d_{j_{4B}}.
\ee
Therefore,
\begin{multline}
\sum_{j_{12}, j_{13}, j_{14}} \left[ \sum_{\substack{j_{1A}, j_{2A}\\ j_{3A}, j_{4A}}} \left[\prod_{i=1}^4 d_{j_{iA}}\right]
\begin{array}{c} \includegraphics[scale=0.4]{10jA.eps} \end{array} \right] \left[ \sum_{j_{1B}} d_{j_{1B}} \begin{array}{c} \includegraphics[scale=0.4]{10jB.eps} \end{array} \right] \\
= \left[d_{j_{23}}\,d_{j_{24}}\,d_{j_{34}}\right]^2\ d_{j_{2B}}\,d_{j_{3B}}\,d_{j_{4B}}.
\end{multline}
One can then further specialize the values of the remaining free spins.

To conclude this Section, we compare briefly our calculation with the 4-dipole in the $\SU(2)$ BF case. Instead of 10j-symbols, the vertex weight is a 15j-symbol (one additional degree of freedom per tetrahedron). As this is a topological case, three delta functions are redundant in the group integral formulation, exactly like in our calculation. The amplitude can thus be regularized the same way, by fixing the spins on three internal faces. Once all delta functions are taken into account, the special solutions \eqref{special conf1} holds at the vertices $A$ and $B$. Therefore, there is no integral like in the right hand side of \eqref{recursion1}. This integral is really the remnant of the way the BF theory is modified to get the BC model (i.e. imposing the simplicity constraints in a specific way), which survives even when the amplitude is restricted by hand to the special BF-like solutions \eqref{special conf1}.

\subsubsection{Using the tetrahedral graph}

We consider a piece of triangulation formed by four 4-simplices, labeled $1,2,3,4$, connected to one another. The tetrahedra they share (called internal) are therefore labeled by pairs $(ij)$, $1\leq i<j\leq 4$. The boundary has eight tetrahedra, each 4-simplex contributing to two, denoted $iA, iB$. The two boundary tetrahedra of the simplex $i$ share a triangle labeled $(AiB)$, for $i=1,2,3,4$. The boundary tetrahedra of the simplices $i,j$ share two triangles, one belonging to the tetrahedra of type-$A$ and one to the tetrahedra of type-$B$. We label these triangles $(ijA)$ and $(ijB)$ and notice that they also belong to the internal tetrahedra $(ij)$. The internal structure has four triangles, which are all shared by three 4-simplices, and are therefore labeled $(ijk)$, for $1\leq i<j<k\leq 4$. The triangle $(ijk)$ belongs to the three internal tetrahedra $(ij), (jk), (ik)$.

\begin{figure}
\includegraphics[scale=0.5]{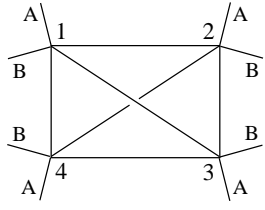}
\caption{ \label{fig:tetrahedral} There are four 4-simplices (represented as vertices), all connected to one another by a tetrahedron (represented as a edge).}
\end{figure}

In the dual picture, 4-simplices are vertices, tetrahedra lines and triangles faces. The 2-complex, denoted $\Gamma_4$, is depicted (as a graph) in the Figure \ref{fig:tetrahedral}. It has four vertices $i=1,2,3,4$, connected to one another by six (internal) lines $(ij)$, for $1\leq i<j\leq 4$. The tetrahedra on the boundary of the gluing are represented by eight half-lines labeled $(iA), (iB)$ for $i=1,2,3,4$, each vertex $i$ having two of them. The external faces are broken faces dual to the boundary triangles. The external face $(AiB)$ goes along the half-lines $(iA)$ and $(iB)$, and there are four of them. There are twelve other external faces, labeled $(ijA)$ (six of them), and $(ijB)$ (six others), for $1\leq i<j\leq 4$. The face $(ijA)$ goes along the half-line $(iA)$, then the internal line $(ij)$ which connects the vertices $i$ to $j$, and continues along the half-line $(jA)$ (similarly for $(ijB)$). The internal faces all have three vertices, which allows to have them labeled $(ijk)$ and there are four of them.

To write the spin foam amplitude on $\Gamma_4$, we fix the spins of the external faces: $j_{i}$ for the face $(AiB)$, $j_{ijA}$ and $j_{ijB}$ for the faces $(ijA), (ijB)$. The internal faces are closed with three lines, and thus carry effective weights of faces with $n=3$. The line $(ij)$ has two group elements $h_{ij}, h_{ji}$, respectively associated to the half-line connected to $i$ and to $j$. The external half-lines have group elements $h_{iA}, h_{iB}$. The partition function is
\begin{multline} \label{Z_Gamma4}
Z_{\Gamma_4}\bigl(\{j_i\}, \{j_{ijA}\}, \{j_{ijB}\}\bigr) \\
= \int \prod_{i=1}^4 dh_{iA} dh_{iB} \prod_{i\neq j} dh_{ij} \prod_{1\leq i<j<k\leq 4} \left[d\gamma_{k(ij)} d\gamma_{j(ki)}\ \delta\bigl( h_{ij} h_{ik}^{-1}\ \gamma_{k(ij)}\,h_{ki} h_{kj}^{-1}\,\gamma_{k(ij)}^{-1}\ \gamma_{j(ki)}\,h_{jk} h_{ji}^{-1}\,\gamma_{j(ki)}^{-1} \bigr)\right]\\
\prod_{i=1}^4 \chi_{j_i}(h_{iA}^{-1} h_{iB})\ \prod_{1\leq i<j\leq 4} \chi_{j_{ijA}}(h_{iA} h_{ij}^{-1})\,\chi_{j_{ijA}}(h_{ji} h_{jA}^{-1})\ \chi_{j_{ijB}}(h_{iB} h_{ij}^{-1})\,\chi_{j_{ijB}}(h_{ji} h_{jB}^{-1}).
\end{multline}
Notice that the internal faces form a spherical bubble, identical to the boundary of a tetrahedron. Therefore, the contribution of this bubble can be evaluated as an application of the result of the Section \ref{sec:single-bubble}. We ignore the external faces (putting their spins to zero), and use the formula \eqref{2dBC} for the BC model on a surface of Euler characteristic $\chi = 2$ with $E=6$ lines, to get
\be
Z_{\Gamma_4}(0,0,0) = \zeta(8) = \frac{\pi^8}{9450}.
\ee
This is obviously finite, meaning that the four deltas in \eqref{Z_Gamma4} are all independent. However, if we can project onto the special configurations \eqref{special conf1}, there would be a gauge symmetry of the BF type, which in the case of $\Gamma_4$ corresponds to one redundant delta (like for any spherical bubble in the BF model). To project onto solutions of the form \eqref{special conf1}, we proceed like in the 4-dipole case. We change some of the characters of the external faces with deltas.

Let us consider
\begin{multline} \label{I_Gamma4}
I_{\Gamma_4} = \int \prod_{i=1}^4 dh_{iA} dh_{iB} \prod_{i\neq j} dh_{ij} \prod_{1\leq i<j<k\leq 4} d\gamma_{k(ij)} d\gamma_{j(ki)}\ \prod_{i=1}^4 \chi_{j_i}(h_{iA}^{-1} h_{iB}) \prod_{1\leq i<j\leq 4} \chi_{j_{ijB}}(h_{iB} h_{ij}^{-1})\,\chi_{j_{ijB}}(h_{ji} h_{jB}^{-1}) \\
\chi_{j_{32A}}(h_{3A} h_{32}^{-1})\,\chi_{j_{41A}}(h_{4A} h_{41}^{-1})\,\chi_{j_{42A}}(h_{4A} h_{42}^{-1}) \biggl[\prod_{i=1,2} \prod_{\substack{j=1,2,3,4\\ j\neq i}} \delta(h_{iA} h_{ij}^{-1})\biggr] \delta( h_{3A} h_{31}^{-1})\ \delta(h_{3A} h_{34}^{-1})\ \delta(h_{4A} h_{43}^{-1})\\
\delta\Bigl( h_{12} h_{13}^{-1}\ \gamma_{3(12)}\,h_{31} h_{32}^{-1}\,\gamma_{3(12)}^{-1}\ \gamma_{2(31)}\,h_{23} h_{21}^{-1}\,\gamma_{2(31)}^{-1} \Bigr)\
\delta\Bigl( h_{13} h_{14}^{-1}\ \gamma_{4(13)}\,h_{41} h_{43}^{-1}\,\gamma_{4(13)}^{-1}\ \gamma_{3(41)}\,h_{34} h_{31}^{-1}\,\gamma_{3(41)}^{-1} \Bigr)\\
\delta\Bigl( h_{12} h_{14}^{-1}\ \gamma_{4(12)}\,h_{41} h_{42}^{-1}\,\gamma_{4(12)}^{-1}\ \gamma_{2(41)}\,h_{24} h_{21}^{-1}\,\gamma_{2(41)}^{-1} \Bigr)
\chi_{j_{234}}\Bigl( h_{23} h_{24}^{-1}\ \gamma_{4(23)}\,h_{42} h_{43}^{-1}\,\gamma_{4(23)}^{-1}\ \gamma_{3(42)}\,h_{34} h_{32}^{-1}\,\gamma_{3(42)}^{-1} \Bigr).
\end{multline}
The group variables we integrate are the same as in \eqref{Z_Gamma4} and the products of these group elements appearing in the integrand are also the same. Only the functions differ. The four external faces $(AiB)$ are untouched, as well as the six external faces $(ijB)$. The characters $\chi_{j_{ijA}}$ along the wedges of the external faces $(ijA)$ have almost all been replaced with deltas, except for the wedge of the face $(23A)$ at the vertex 3, the wedge of the face $(14A)$ at the vertex 4, and the wedge of the face $(24A)$ at the vertex 4. Finally, the delta on the internal face $(234)$ has been changed with a single mode $\chi_{j_{234}}$, to avoid a divergence due to a gauge symmetry as we will see.

Now let us consider the effects of all these new deltas in $I_{\Gamma_4}$. We have $h_{12}=h_{13}$, $h_{21}=h_{23}$. Therefore, the constraint on the face $(123)$ simplifies to $h_{31}=h_{32}$,
\begin{multline}
\delta(h_{1A} h_{12}^{-1})\, \delta(h_{1A} h_{13}^{-1})\ \delta(h_{2A} h_{21}^{-1})\, \delta(h_{2A} h_{23}^{-1})\ \delta\Bigl( h_{12} h_{13}^{-1}\ \gamma_{3(12)}\,h_{31} h_{32}^{-1}\,\gamma_{3(12)}^{-1}\ \gamma_{2(31)}\,h_{23} h_{21}^{-1}\,\gamma_{2(31)}^{-1} \Bigr) \\
= \delta(h_{1A} h_{12}^{-1})\, \delta(h_{1A} h_{13}^{-1})\ \delta(h_{2A} h_{21}^{-1})\, \delta(h_{2A} h_{23}^{-1})\ \delta(h_{31} h_{32}^{-1}).
\end{multline}
Similarly, we have $h_{13}=h_{14}, h_{31}=h_{34}$ which means that the constraint on the face $(134)$ simplifies to $h_{41}=h_{43}$,
\begin{multline}
\delta(h_{1A} h_{13}^{-1})\, \delta(h_{1A} h_{14}^{-1})\ \delta(h_{3A} h_{31}^{-1})\, \delta(h_{3A} h_{34}^{-1})\ \delta\Bigl( h_{13} h_{14}^{-1}\ \gamma_{4(13)}\,h_{41} h_{43}^{-1}\,\gamma_{4(13)}^{-1}\ \gamma_{3(41)}\,h_{34} h_{31}^{-1}\,\gamma_{3(41)}^{-1} \Bigr)\\
= \delta(h_{1A} h_{13}^{-1})\, \delta(h_{1A} h_{14}^{-1})\ \delta(h_{3A} h_{31}^{-1})\, \delta(h_{3A} h_{34}^{-1})\ \delta(h_{41} h_{43}^{-1}),
\end{multline}
and for the face $(124)$, we get $h_{42}=h_{41}$,
\begin{multline}
\delta(h_{1A} h_{12}^{-1})\,\delta(h_{1A} h_{14}^{-1})\ \delta(h_{2A} h_{21}^{-1})\,\delta(h_{2A} h_{24}^{-1})\ \delta\Bigl( h_{12} h_{14}^{-1}\ \gamma_{4(12)}\,h_{41} h_{42}^{-1}\,\gamma_{4(12)}^{-1}\ \gamma_{2(41)}\,h_{24} h_{21}^{-1}\,\gamma_{2(41)}^{-1} \Bigr)\\
= \delta(h_{1A} h_{12}^{-1})\,\delta(h_{1A} h_{14}^{-1})\ \delta(h_{2A} h_{21}^{-1})\,\delta(h_{2A} h_{24}^{-1})\ \delta(h_{41} h_{42}^{-1}).
\end{multline}
As a result of all the deltas in \eqref{I_Gamma4}, we find the set of solutions of the contraints,
\be
\cF_{\Gamma_4} = \left\{ h_{12}=h_{13}=h_{14}=h_{1A}, h_{21}=h_{23}=h_{24}=h_{2A}, h_{31}=h_{32}=h_{34}=h_{3A}, h_{41}=h_{42}=h_{43}=h_{4A} \right\}.
\ee
and the $\gamma$s can take arbitrary values. The character on the fourth face, $(234)$, is thus simply evaluated on the identity,
\be
\chi_{j_{234}}\Bigl( h_{23} h_{24}^{-1}\ \gamma_{4(23)}\,h_{42} h_{43}^{-1}\,\gamma_{4(23)}^{-1}\ \gamma_{3(42)}\,h_{34} h_{32}^{-1}\,\gamma_{3(42)}^{-1} \Bigr)_{|\cF_{\Gamma_4}} = d_{j_{234}}.
\ee
Clearly, a delta on that face would have been redundant, and caused a divergence of the type $\delta(\mathbbm{I}) = \sum_{j_{234}} d_{j_{234}}^2$. This is the sign of the gauge symmetry which exists when projecting onto $\cF_{\Gamma_4}$.

The other characters going along the half-lines $(iA)$ simplify,
\be
\chi_{j_{32A}}(h_{3A} h_{32}^{-1})\,\chi_{j_{41A}}(h_{4A} h_{41}^{-1})\,\chi_{j_{42A}}(h_{4A} h_{42}^{-1})_{|\cF_{\Gamma_4}} = d_{j_{32A}}\ d_{j_{41A}}\ d_{j_{42A}}.
\ee
The only remaining non-trivial part is the integrals over $h_{iB}$. Performing the changes of variables $h_i \equiv h_{iB} h_{ij}^{-1}$, it finally becomes
\be
I_{\Gamma_4} = d_{j_{234}}\ d_{j_{32A}}\ d_{j_{41A}}\ d_{j_{42A}} \prod_{i=1}^4 \left[\int dh_i\ \chi_{j_i}(h_i)\,\prod_{j\neq i} \chi_{j_{ijB}}(h_i)\right],
\ee
where we recognize these integrals as the same as the ones on the right hand side of \eqref{recursion1}.

A second way to evaluate $I_{\Gamma_4}$ is by expanding all deltas as $\delta=\sum_j d_j \chi_j$ and integrating the group elements $\gamma$ using the orthogonality relation \eqref{orthogonality}. For each vertex $i=1,2,3,4$, we get a 10j-symbol,
\be
\int dh_{iA}\,dh_{iB}\,\prod_{j\neq i} dh_{ij}\ \chi_{j_i}(h_{iA} h_{iB}^{-1}) \prod_{j\neq i} \chi_{j_{ijA}}(h_{iA} h_{ij}^{-1})\,\chi_{j_{ijB}}(h_{iB} h_{ij}^{-1}) \prod_{\substack{j<k \\j,k\neq i}} \chi_{j_{ijk}}(h_{ij} h_{ik}^{-1}).
\ee
Equating the two ways to evaluate $I_{\Gamma_4}$ finally gives
\begin{multline}
d_{j_{234}}\ d_{j_{32A}}\ d_{j_{41A}}\ d_{j_{42A}} \prod_{i=1}^4 \left[\int dh_i\ \chi_{j_i}(h_i)\,\prod_{j\neq i} \chi_{j_{ijB}}(h_i)\right] = \frac{1}{d_{j_{234}}^2}
\sum_{j_{123}, j_{134}, j_{124}} \frac{1}{d_{j_{123}}\,d_{j_{124}}\,d_{j_{134}}} \\
\left[\sum_{\substack{j_{12A}, j_{13A}\\ j_{14A}}} d_{j_{12A}} d_{j_{13A}} d_{j_{14A}} \begin{array}{c} \includegraphics[scale=0.4]{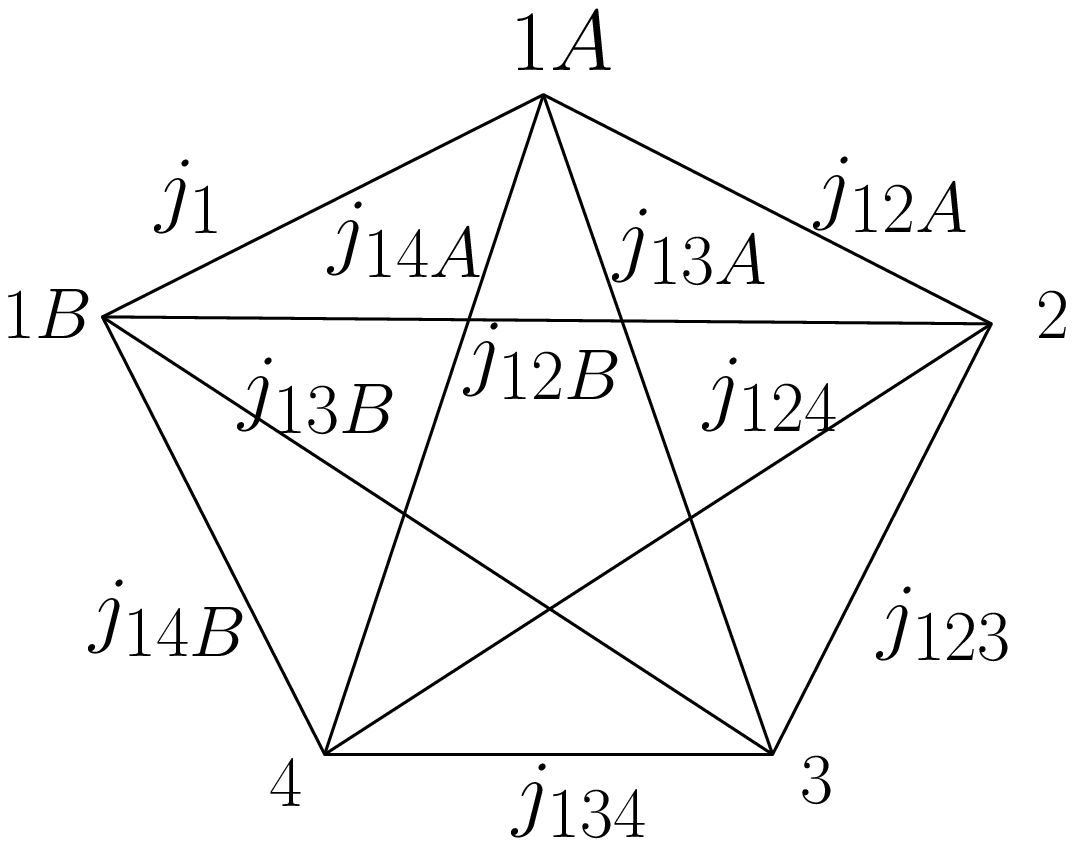} \end{array} \right]
\left[\sum_{\substack{j_{21A}, j_{23A}\\ j_{24A}}} d_{j_{21A}} d_{j_{23A}} d_{j_{24A}} \begin{array}{c} \includegraphics[scale=0.4]{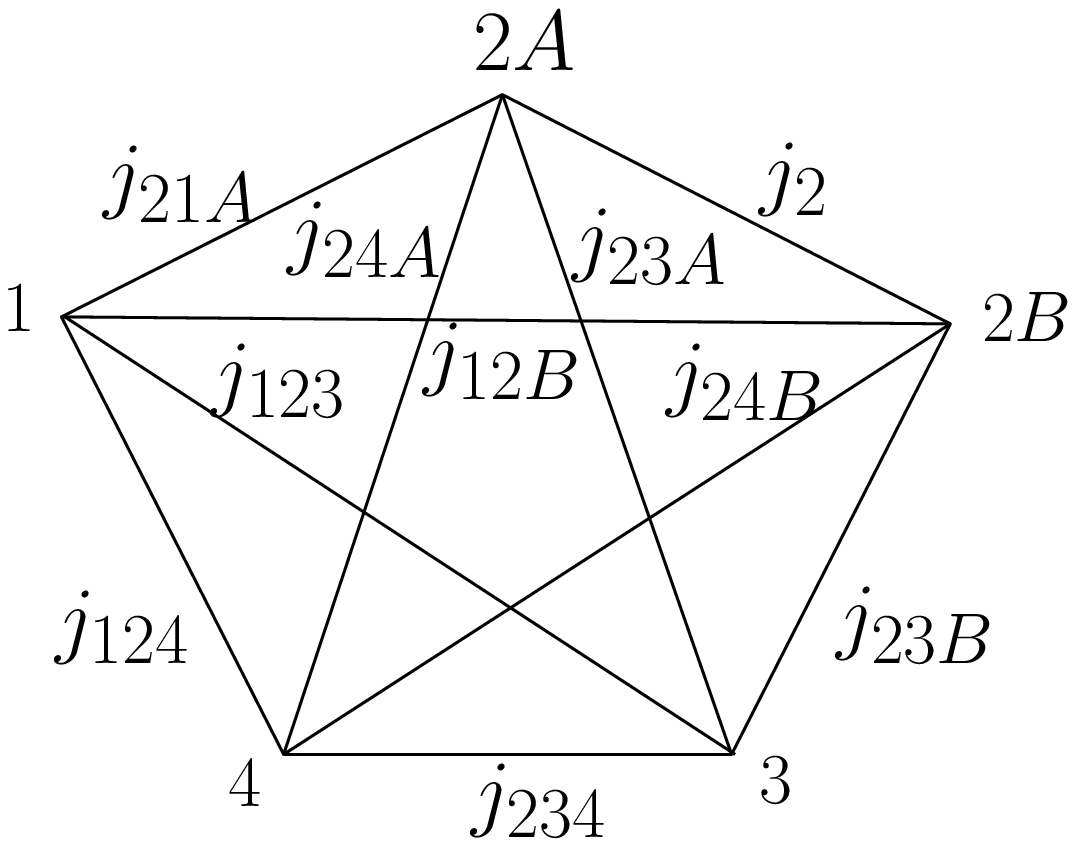} \end{array} \right] \\
\left[\sum_{j_{31A}, j_{34A}} d_{j_{31A}} d_{j_{34A}} \begin{array}{c} \includegraphics[scale=0.4]{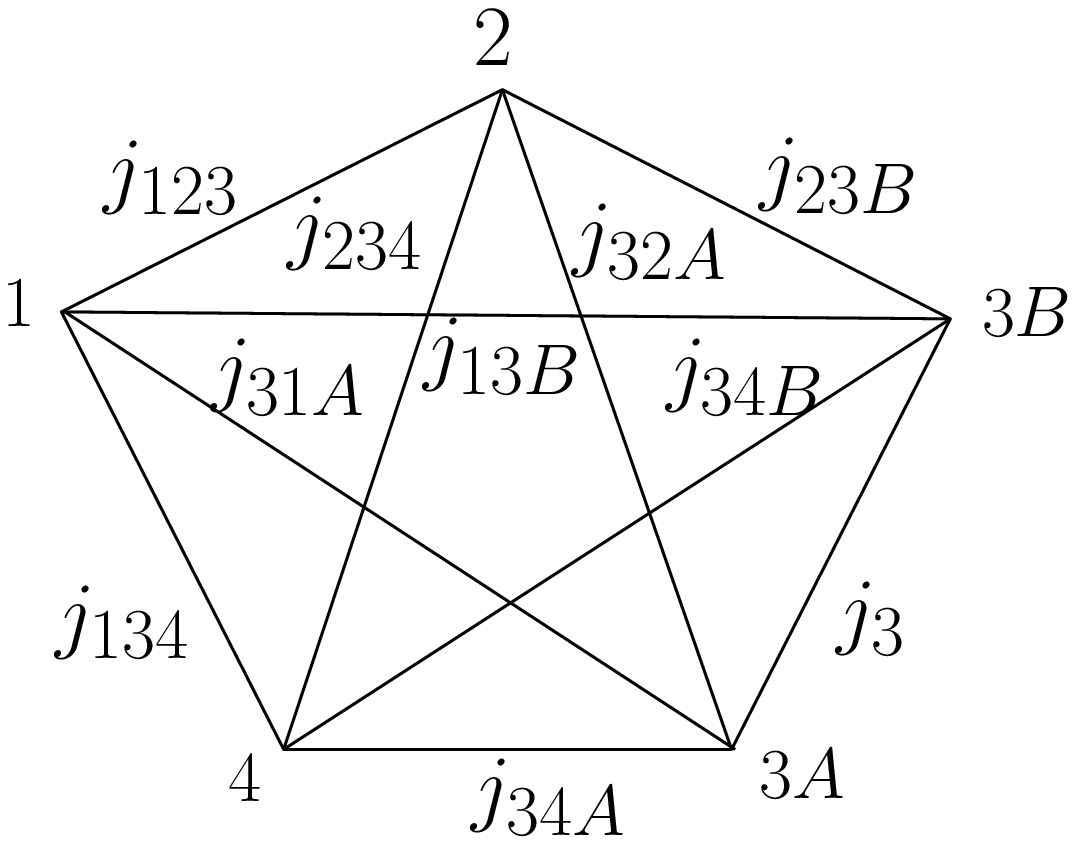} \end{array} \right]
\left[\sum_{j_{43A}} d_{j_{43A}} \begin{array}{c} \includegraphics[scale=0.4]{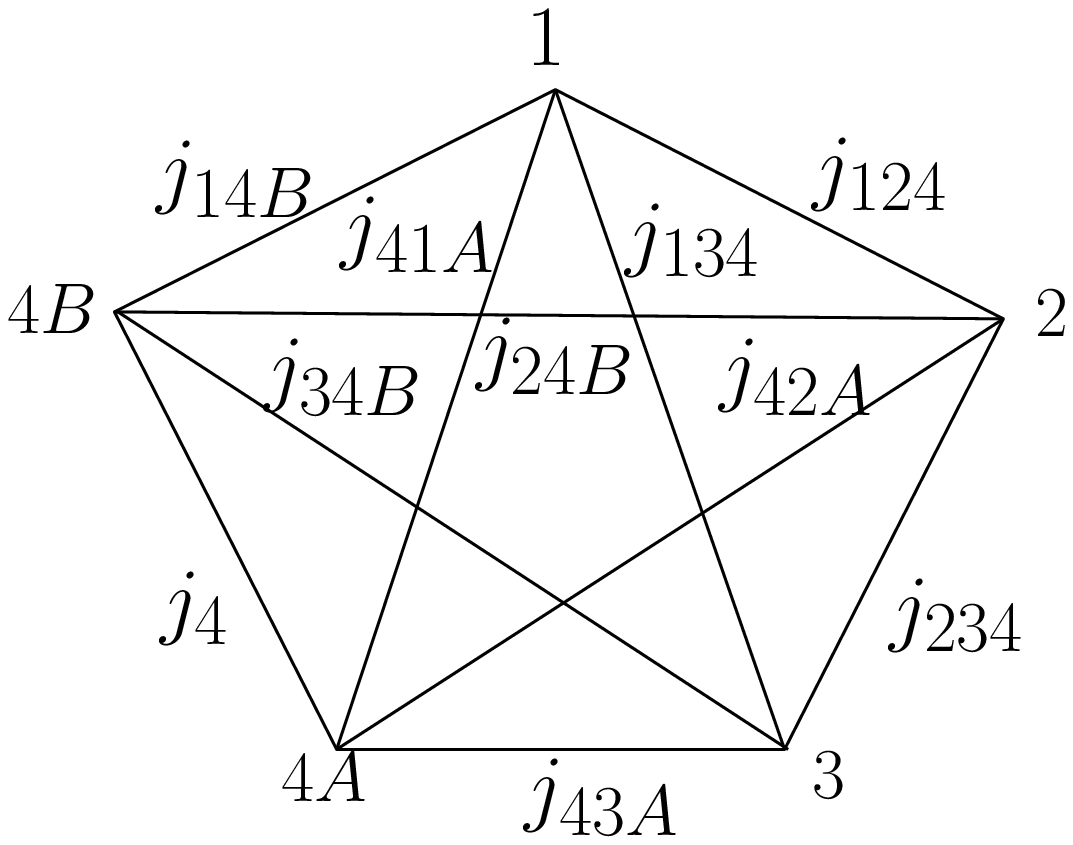} \end{array} \right]
\end{multline}

In summary, as in the case of classical discrete gravity, there are special solutions to the delta function constraints describing the BC model around which gauge symmetries occur. By changing the partition functions appropriately we can enforce a projection onto these special solutions. These altered partition functions will have gauge symmetries, and this can be used to derive equations involving the vertex amplitude of the BC model.

\section{Discussion}
\label{discussion}

We discussed possible divergences in spin foam models, in particular in the Barrett-Crane model. The reason for considering this model is that the relation of divergences to gauge symmetries is easily found in such a model, for which the partition function can be rewritten as an integral over a space of flat connections. This also explains our choice of edge weight factors and face weights.

We presented a simple method to  determine the (single bubble) divergences for general spin foam models. We argue that to this end one just needs to evaluate the spin foam model on a two--dimensional surface and we gave an explicit formula for a large class of models, encompassing BC and EPRL/FK. This allows to deduce the influence of the edge and face weights on the convergence of the models very easily. We noticed that requiring invariance under face divisions and in addition under (two--valent) edge subdivisions leads to triangulation invariant 2D models, which have however divergent spherical bubbles.

We discussed in detail the effective face weights as they capture the basic, possibly distributional, building blocks for the models. We found that for the BC model the effective face weights (for faces with more than two edges) are (almost everywhere) regular functions. For the EPRL/FK models, finiteness depends on the choice of edge and face weight factors. However applying our arguments on how to evaluate spherical bubbles, one can consider the square of the effective face weights. This will identify a distributional character of the effective face weights also for these models. The (Lorentzian) EPRL/FK model has been argued to be finite for the 1-5 move and have only a logarithmic divergence for the 4-dipole \cite{riello}. This suggest that the effective face weights for faces with more than two edges might also be finite functions and not be distributional in this case.

For spin foams recent work \cite{fw} suggested the notion of wave front sets, which specifies the non--smooth part of a distribution, in order to study the large spin limit and regularization issues. The wave front sets have been identified for both the Barrett--Crane and the EPRL model in \cite{fw}, thus there is clearly the potential for divergent behavior. Indeed if the wave front sets would correspond to divergences, regularization is needed (to define products of distributions). In that case, the conclusions of \cite{fw} for the large spin limit regime would have to be reconsidered, as this work assumed that products can be formed from the distributions that occur as amplitudes in spin foams. Here we found that (for higher than two--valent faces) the wave front sets for BC correspond to non--continuous or non--smooth but also non--divergent behavior (modulo sets of measure zero).



Furthermore we analyzed in detail the 4-dipole case for the BC model. With our choice of edge and face weight the spin picture does not allow for a definite conclusion regarding finiteness. We therefore used the fact that the BC amplitudes can be rewritten as integrals over some sets of flat connections, which renders the problem accessible through the method of \cite{homology-bf}. We showed that no redundancies of delta functions arise, thus excluding divergences due to this reason (another possible source are the measure zero singularities though).

Redundancies of delta functions would be the sign of gauge symmetries. As such redundancies are not occurring -- and in addition we found convincing arguments that the BC model is finite on a regular (i.e. involving only faces with more than two faces) triangulation --  we have to conclude that gauge symmetries, which could be connected to diffeomorphisms, are not present. This even seems to hold for configurations, for instance the 5-1 Pachner move, for which the symmetries exist on the classical (Regge) level \cite{measure-sebastian}.

There are however special solutions (of measure zero) for which delta function redundancies can be identified. Similarly there are measure zero solutions in gravity (the flat solutions) around which (linearized) gauge symmetries can be found.
We discussed those special solutions and described the related gauge symmetries for the BC model. We developed a method to derive associated Ward-identity-like equations on the vertex amplitude. This is the first proposal which enables to extract constraints from a quantum theory with broken gauge symmetries, i.e. equations that have to hold for the boundary wave function (which modulo measure factors can be identified with the vertex amplitude), extending this way the tools introduced \cite{recursion} for topological theories.

The question arises whether those special symmetries can be the seed for the occurrence of more general symmetries, that might emerge under coarse graining \cite{eckert}. The heuristic argument is that coarse graining leads to an effective description of the coarse model on a much finer triangulation. On this fine triangulation the curvature per building block is very small, so that one is near the flat case, that is on the special solutions around which gauge symmetries do exist.

This mechanism actually works for classical systems \cite{perfect-actions} as well as for 1D quantum systems \cite{perfectdis-pathint}. In this case the amplitudes might become more and more divergent under coarse graining, as is indeed the case for 1D discretized quantum systems \cite{perfectdis-pathint}. We will leave this question for future work.

Another interesting question is whether the wave front analysis performed in \cite{fw} can be used to analyze the gauge symmetries around special solutions also for more general spin foam models. Wave fronts are a refinement of the singular support to the co--tangent space. We also used the co--tangent space in (\ref{71}) to define the notion of gauge symmetries we applied in this work. This might make the methods presented here applicable to other spin foam models as well.

\appendix
\section{$\SU(2)$ calculus} \label{app:su2calculus}

We parametrize group elements as $h = e^{i\theta\,\hat{n}\cdot\vec{\sigma}} = \cos\theta\, \mathbbm{I} +i\,\sin\theta\,\hat{n}\cdot\vec{\sigma}$, where $\hat{n}\in S^2$ is the rotation axis and $\theta\in[0,\pi]$ is the class angle. The vector $\vec{\sigma}=(\sigma_x,\sigma_y,\sigma_z)$ is the 3-vector formed by the Pauli matrices\footnote{They read
\be
\sigma_x = \begin{pmatrix} 0&1\\1&0\end{pmatrix},\quad \sigma_y = \begin{pmatrix} 0&-i\\i&0\end{pmatrix},\quad \sigma_x = \begin{pmatrix} 1&0\\0&-1\end{pmatrix}.
\ee}, which transforms as a co-vector under the adjoint action,
\be
g\,\sigma_i g^{-1} = \sum_{j=x,y,z} R(g^{-1})_{ij} \sigma_j,
\ee
$R(g)$ being the rotation matrix in the vector representation, $g\in\SU(2)$. From this, the orbit of the adjoint action on the group is found,
\be
g\,h\,g^{-1} = \cos \theta\,\mathbbm{I} +i\,\sin\theta\,\bigl(R(g) \hat{n}\bigr)\cdot \vec{\sigma} = e^{i\theta\,(R(g)\hat{n})\cdot\vec{\sigma}},
\ee
meaning that $g$ rotates the rotation axis of $h$ without changing its class angle.

The matrix elements in the irreducible representation of spin $j\in\N/2$ satisfy the orthogonality relation
\be \label{orthogonality}
\int_{\SU(2)} dh\ D^{(j_1)}_{m_1n_1}(h)\,\overline{D^{(j_2)}_{m_2 n_2}(h)} = \frac1{d_{j_1}}\,\delta_{j_1 j_2}\ \delta_{m_1 m_2}\,\delta_{n_1 n_2}.
\ee
Here $dh$ is the normalized Haar measure, $d_j\equiv 2j+1$ is the dimension of the representation and $D^{(j)}$ the Wigner matrices. A useful property is $D^{(j)}_{mn}(h^{-1}) = \overline{D^{(j)}_{nm}(h)}$ for any $h\in \SU(2)$. The character in the representation of spin $j$ is the trace,
\be
\chi_j(h) = \chi_j(h^{-1}) = \sum_{m=-j}^j e^{im\theta} = \frac{\sin d_j \theta}{\sin\theta}.
\ee
It satisfies $\chi_j(\mathbbm{I}) = d_j$. The convolution of characters is
\be \label{convolution}
\int dg\ \chi_{j_1}(h_1^{-1} g)\ \chi_{j_2}(g^{-1} h_2) = \delta_{j_1 j_2}\ \frac1{d_{j_1}}\,\chi_{j_1}(h_1^{-1} h_2).
\ee

Functions in $L^2(\SU(2),dh)$ admit expansions over the Wigner matrices, $f(h) = \sum_{j\in\N/2}\sum_{m,n=-j}^j \sqrt{d_j}\,f^{(j)}_{mn} D^{(j)}_{mn}(h)$, which is the Fourier expansion. Class functions are the functions invariant under the adjoint action, so that they only depend on the conjugacy class, i.e. the class angle. Characters provide a basis of class functions. The Dirac delta over $\SU(2)$ is the distribution such that $\int_{\SU(2)} dg\,\delta(g)\,f(g) = f(\mathbbm{I})$ and it has the expansion
\be
\delta(g) = \sum_{j\in\N/2} d_j\,\chi_j(g).
\ee
The delta over the conjugacy class of angle $\psi$ is
\be
\delta_\psi(h) = \int_{\SU(2)} d\gamma\ \delta\bigl(h\,\gamma\,g_\psi\,\gamma^{-1}\bigr) = \sum_{j\in\N/2} \chi_j(g_\psi)\ \chi_j(h),
\ee
where $g_\psi \in\SU(2)$ is any representative of the conjugacy class.


\section*{\small Acknowledgements}

It is a pleasure to thank Wojciech Kaminski for extensive discussions.
Research at Perimeter Institute is supported by the Government of Canada through Industry Canada and by the Province of Ontario through the Ministry of Research and Innovation.


\end{document}